\documentclass[fleqn,usenatbib]{mnras}

\usepackage{newtxtext,newtxmath}

\usepackage[T1]{fontenc}
\usepackage{ae,aecompl}
\usepackage{soul}

\usepackage{graphicx}	
\usepackage{amsmath}	
\usepackage{amssymb}	

\title[\texttt{Clusterix} 2.0]{\texttt{Clusterix} 2.0. A Virtual Observatory tool to estimate cluster membership probability}

\author[Balaguer-N\'u\~nez et al.]{
L. Balaguer-N\'u\~nez$^{1}$\thanks{e-mail: lbalaguer@fqa.ub.edu},
M. Lopez$^{2}$,
E. Solano$^{2}$,
D. Galad\'{\i}-Enr\'{\i}quez$^{3}$,
C. Jordi$^{1}$,
\newauthor
F. Jimenez-Esteban$^{2}$,
E. Masana$^{1}$,
J. Carbajo-Hijarrubia$^{1}$
and E. Paunzen$^{4}$
\\
$^{1}$Institut de Ci\`encies del Cosmos, Universitat de Barcelona (IEEC-UB), Mart\'i i Franqu\`es 1, E-08028 Barcelona, Spain  \\ 
$^{2}$Departmento de Astrof\'{\i}sica, Centro de Astrobiolog\'{\i}a (CSIC-INTA), ESAC Campus, Camino Bajo del Castillo s/n,\\
E-28692 Villanueva de la Ca\~nada, Madrid, Spain; Spanish Virtual Observatory \\
$^{3}$Observatorio de Calar Alto, Sierra de los Filabres, E-04550-G\'ergal (Almer\'{\i}a), Spain \\
$^{4}$Departament of Theoretical Physics and Astrophysics. Marsaryk University. Brno, Czech Republic \\
}

\date{Accepted XXX. Received YYY; in original form ZZZ}

\pubyear{2019}

\begin{document}
\label{firstpage}
\pagerange{\pageref{firstpage}--\pageref{lastpage}}
\maketitle

\begin{abstract}
\texttt{Clusterix} 2.0 is a web-based, Virtual Observatory-compliant, interactive tool for the determination of membership probabilities in  stellar clusters based on proper motion data using a fully non-parametric  method. In the area occupied by the cluster, the frequency function is made up of two contributions: cluster and field stars. The tool performs an empirical determination of the frequency functions from the Vector-Point Diagram without relying in any previous assumption about their profiles. \texttt{Clusterix} 2.0 allows to search in an interactive way the appropriate spatial areas until an optimal separation of the two populations is obtained. Several parameters can be adjusted to make the calculation computationally feasible without interfering in the quality of the results. The system offers the possibility to query different catalogues, such as {\em Gaia}, or upload the user own data. The results of the membership determination can be sent via SAMP to VO tools like TopCat.

We apply \texttt{Clusterix} 2.0 to several open clusters with different properties and environments to show the capabilities of the tool: an area of five degrees around NGC~2682 (M~67), an old, well known cluster; a young cluster NGC~2516 with a striking elongate structure extended up to four degrees; NGC~1750 \& NGC~1758, a pair of partly overlapping clusters; in the area of NGC~1817 we confirm a little-known cluster, Juchert~23; and in an area with many clusters we disentangle the existence of two overlapping clusters where only one was previously known: Ruprecht~26 and the new, Clusterix~1.

\end{abstract}
 



\section{Introduction}
\label{intro}
Open clusters (OCs) are coeval groups of stars formed from the same molecular cloud and, thus, having the same age and initial chemical composition. This makes them ideal targets to study the formation and evolution of stellar objects. Open clusters are also valuable tools for undertaking dynamic and kinematic studies of our Galaxy. For instance, clusters have been used to determine the spiral structure of the Galaxy and to investigate the dynamics and the chemical evolution of the Galactic disk \citep{Frinchaboy2008}. 
They are good tracers to follow the metallicity gradient of the Milky Way \citep{Carrera2011, Netopil2016, Casamiquela2017} and its evolution through time \citep{Magrini2009, Jacobson2016}, providing insight  on the formation of the Galactic disk. Studies of the kinematics of OCs and reconstructions of their individual orbits \citep{Wu2009, tristan2016} help us to understand the internal processes of heating \citep{Quillen2018} and radial migration \citep{Anders2017}, and how they affect the chemodynamical evolution of the disk. Some very perturbed orbits might also provide evidence for recent merger events and traces of past accretion from outside the Galaxy \citep{tristan2016}.

 Open clusters are not only useful tracers of the Milky Way structure but are also interesting targets in their own right. Although most stars in the Milky Way are observed in isolation, it is believed that most (possibly all) stars form in clustered environments and spend at least a short amount of time gravitationally bound with their siblings \citep{Clarke2000, portegieszwart2010ARA&A}, embedded in their progenitor molecular cloud. A majority of such systems will be disrupted in their first few million years of existence, due to mechanisms possibly involving gas loss driven by stellar feedback \citep{Brinkmann2017} or encounters with giant molecular clouds \citep{Gieles2006}. Nonetheless, a fraction will survive the embedded phase and remain bound over longer timescales.

{\em Gaia} Data Release 2 (DR2), published on April 2018, provides a 5-parameter astrometric solution (celestial position, proper motions in right ascension and declination, and parallax) and magnitudes in three photometric filters ($G$, $G_{\rm BP}$, and $G_{\rm RP}$) for more than 1.3 billion sources with unprecedented precision and accuracy \citep{Brown2018}. {\em Gaia} is a fundamental resource to study the known OCs and to discover new ones, as well as to discard clusters previously identified \citep{tristan2018b, tristan2019, CastroGinard2018}. With this aim, an accurate method to calculate cluster membership is essential in order to conduct further studies \citep{Frinchaboy2008}.

The determination of the mean properties of open clusters (like radial velocity or metallicity) requires prior knowledge of their member stars to optimise the costly process of obtaining and reducing high resolution spectroscopic data on a large scale. Some information is already available mainly for radial velocity \citep{soubiran2018, Carrera2019a}. Moreover, knowing the membership probability of the stars in the cluster area helps to select member stars on photometric diagrams, to determine the age of the cluster by isochrone fitting \citep[e.g.][]{Bossini2019}. Hence, a precise identification of the stars that compose a cluster is critical to accurately determine the kinematic and fundamental parameters of the clusters (age, total mass, etc.), which are essential for studies of Galactic dynamics. 

In order to efficiently exploit the wealth and quality of {\em Gaia} data, a variety of new approaches and new tools must be considered. Several automatic approaches have been developed to study the reality of known OCs \citep{tristan2018a} and to discover new ones \citep{CastroGinard2018, tristan2018b}. In our case, we aim to develop a web tool to facilitate membership studies from proper motions data to any user that requires a tailor-made study on any specific data set. 

\texttt{Clusterix} 1.0 \citep{Sezima2015} was a web tool based on the implementation of the non-parametric method for membership segregation \citep{Galadi1998, BalaguerNunez2007} developed at the Universitat de Barcelona (Spain) in collaboration with the Masaryk University (Czech Republic) as a complement to the WEBDA\footnote{http://webda.physics.muni.cz} database \citep{Netopil2012} of observational data on stars in open clusters.
In this paper we present \texttt{Clusterix} 2.0\footnote{http://clusterix.cab.inta-csic.es}, an upgraded, much more powerful, VO-compliant version of the web tool jointly developed by the Universitat de Barcelona and the Spanish Virtual Observatory\footnote{http://svo.cab.inta-csic.es } \citep{clusterix2017}. Compliance with the Virtual Observatory is key for an optimum gathering from VO services and catalogues of additional information (as parallaxes, or radial velocities) than can be used to refine the membership determination. Moreover, stellar physical parameters as effective temperatures, radii, luminosities or metallicities can be estimated from VO services like VOSA\footnote{http://svo2.cab.inta-csic.es/theory/vosa/} \citep{Bayo2008}. 

The paper is organized as follows: In Sect.~\ref{formalism} we present the formalism, Sect.~\ref{workflow} describes the implementation, while Sect.~\ref{science} presents different scientific cases to show the capabilities of the tool. Finally, Sect.~\ref{conclusions} contains the conclusions. 

\section{The formalism: membership probabilities from proper motions}
\label{formalism}
Kinematic segregation of stellar cluster members relies on the fact that the cluster displays a common spatial motion that distinguishes it from the field population. The kinematic segregation may relay on radial velocities, on proper motions, or on both.

The classical approach to cluster/field segregation from proper motion data has been traditionally the parametric method. This method assumes the existence of two overlapping populations in the 2-D proper motion vector-point diagram (VPD): {\it cluster} and {\it field}. The corresponding frequency functions are modelled as parametric Gaussian functions: a circular Gaussian model is adopted for the cluster distribution, while a bivariate (elliptical) Gaussian describes the field. Membership probabilities are later derived from the fits (see \cite{Sanders1971} for a full description of the methodology). However, the parametric approach shows several drawbacks, mainly related to the fact that this Gaussian modelling of the frequency functions is not always realistic, and these disadvantages are not fully overcome by the several improvements introduced into this approach by different authors during the last decades (see a description of a quite sophisticated implementation of parametric methods, for instance, in \citealt{ZhaoHe1990}). More recently, this approach has been highly improved to treat multidimensional data (photometric, astrometric) \citep{Sarro2014} and even treating missing values \citep{Olivares2018}. 

The cluster/field segregation from proper motions can also be approached by means of non-parametric methods, aimed to empirically model the {\it cluster} and {\it field} proper motion distributions that build up the particular VPD of the stars in the sky area under study. Clustering algorithms have been applied to the assignment of membership probabilities in OCs from multidimensional data. UPMASK \citep{KroneMartins2014} was developed to use only photometry and positions, and has been afterwards adapted to {\em Gaia} data \citep{tristan2018a,tristan2018b} based only on proper motions and parallaxes. A similar approach using DBSCAN and machine learning algorithms \citep{CastroGinard2018,CastroGinard2019} has conducted to discover new clusters from an automatic search, demonstrating that still many nearby cluster remain to be found \citep{tristan2019}. And a 
complex scenario with a huge amount of new moving groups \citep{Faherty2018} and star forming regions \citep{Prisinzano2018} is beginning to appear. 

There are as well other authors that mixed both approaches in different ways \citep[see e.g.][]{Sampedro2017}. Although, on average, all methods yield similar results, there are also specific differences between them depending in the type of data to be used or in some particular clusters.

\subsection{The non-parametric approach as implemented in \texttt{Clusterix} 2.0}
\label{nonparametric}
 The non-parametric method implemented in \texttt{Clusterix} 2.0 follows a philosophy explained in length in \cite{BalaguerNunez2004, BalaguerNunez2005, BalaguerNunez2007, Galadi1998b}. However, several details are specific to \texttt{Clusterix} 2.0 and, for the sake of clarity, we will formalise the algorithm in the following paragraphs. 

\subsubsection{Assumptions}
\label{assumptions}
\texttt{Clusterix} 2.0 performs an empirical determination of the frequency functions without relying on any previous assumption about their profiles. First of all, we admit that some sky area has been selected as {\it work space} (we will use label 't' to refer to this area), and that celestial coordinates and proper motions are available for the stars in that region. The procedure relies on several further assumptions:

\begin{enumerate}
\item The population of field stars is spatially and kinematically uniform over the work space.
\item There is a non-field population that outstands, from the spatial density point of view, as an excess in the surface density of stars.
\item The non-field population does not occupy all the work space, but is spatially concentrated allowing to distinguish two regions in the work space: the {\it only field} (label 'f') region, dominated by field stars, and the {\it cluster+field} (label 'c+f') region, that includes both field and non-field stars.
\item Field stars and non-field stars show proper motion distributions on the VPD that are distinctive.
\end{enumerate}

The second assumption implies that, at least in its current shape, \texttt{Clusterix} 2.0 yields accurate estimations about the expected number of cluster members for systems that show some spatial contrast against the background of field stars in the work space (see Eqs. \ref{expectedmembers} and \ref{scalingfactor}). This contrast helps, too, in the selection of the field and cluster regions in the work space, at least in the very first steps of the process. 

It is worth noting that these assumptions do not include any consideration about the shape of the proper motion distributions, that may be highly asymmetric. In particular, we underline that the non-field population may be, in principle, composed from more than one cluster, for instance. In fact, two of the examples of scientific exploitation included in this paper (see Sects.~\ref{NGC1750} and \ref{Rup26}) clearly illustrate this situation. 

\subsubsection{Observables and scaling factors}
\label{observables}

Once the three regions, t, f and c+f, have been defined, the first step leads to two observables for each of them: the spatial areas ($A_{\textnormal t}$, $A_{\textnormal f}$, $A_{\textnormal {c+f}}$) and the number of stars they contain ($n_{\textnormal t}$, $n_{\textnormal f}$, $n_{\textnormal {c+f}}$). 

The first quantity to be derived from these observables is the expected number of non-field stars in the work space $A_{\textnormal t}$, $N_{\textnormal c}$:

\begin{equation}
\label{expectedmembers}
N_{\textnormal c}=n_{\textnormal t}-\frac{A_{\textnormal t}}{A_{\textnormal f}}n_{\textnormal f}
\end{equation}

In Eq. \ref{expectedmembers} it is implicit the computation of the expected total number of field stars in the work space, $N_{\textnormal f}$, computed from $n_{\textnormal f}$ through a straightforward scaling factor, and relying on assumptions (i) and (ii):

\begin{equation}
\label{scalingfactor}
N_{\textnormal f}=\frac{A_{\textnormal t}}{A_{\textnormal f}}n_{\textnormal f} 
\end{equation}

The areas $A_{\textnormal t}$, $A_{\textnormal f}$ and $A_{\textnormal {c+f}}$ will act several times as scaling factors in later steps of the procedure. 

\subsubsection{Raw, normalised and scaled frequency functions}
\label{frequency}

Let us consider the VPD for an arbitrary set of $N$ stars. The {\em frequency function} would provide the density of stars in this diagram, in units of stars per unit area. If, for instance, our proper motions are in units of milliarcseconds per year (mas yr$^{-1}$), the frequency function would be measured in stars per (mas yr$^{-1}$)$^2$. Parametric methods try to fit functions to reproduce the observed density distribution of stars in the VPD, but our non-parametric approach measures the true local density in the real diagram and provides the empirical distribution in the form of a table. 

To this end, the VPD is divided into a (large) number of square cells. Each cell is described by its position in the array according to its row and column numbers, $(i,j)$, and by the proper motion values corresponding to its centre, $(a_i,b_j)$. An empirical value of the frequency function is assigned to each cell, computed as follows:

\begin{equation}
\label{ffdef}
\psi(a_i,b_j)=\sum_{k=1}^{N} K(a_k,b_k) 
\end{equation}

Here, $(a_k,b_k)$ are the coordinates (proper motions in our case) of the $k$-th point of the observed sample with a total of $N$ elements, and $K$ is the so-called {\em kernel function} \citep{Galadi1998}, a function that assigns to each element a weight that decreases with the distance from the centre of cell $(i,j)$. We use normal Gaussian kernel functions defined as:

\begin{equation}
\label{kernel}
K(a,b)=\frac{1}{2\pi h^2} \exp \left[ -\frac{1}{2} \frac{(a-a_i)^2+(b-b_i)^2}{h^2}\right]
\end{equation}

This way, a point located exactly at the centre of cell $(i,j)$ provides the maximum contribution to compute the local density, and all other points in the VPD contribute progressively less to the density evaluated at that cell. Number $h$ is usually named in this context {\em the smoothing parameter}, and it is measured in the same units as the proper motions. 
\texttt{Clusterix} 2.0 allows the user to modify $h$ at will, although the system proposes by default a value computed by means of Silverman's rule \citep{Silverman1986}:

\begin{equation}
\label{silverman}
 h = \left( \frac{4}{d+2}\right)^{1/(d+4)}\sigma N^{-1/(d+4)}
\end{equation}

In Eq. \ref{silverman} $N$ is, again, the number of elements in the VPD that is being analysed, $\sigma$ is the average marginal variance of the sample and $d$ is the dimension of the space (in this case, $d=2$).

The previous formalism is applied in \texttt{Clusterix} 2.0 to the VPDs observed at the regions c+f and f. The same value of $h$ is used in both computations, and the default value according to Eq.~\ref{silverman} is deduced from the sample in c+f region. This way, the empirical, raw frequency functions are computed for the mixed (c+f) and only field (f) samples: $\psi_{\textnormal {c+f}}$ and $\psi_{\textnormal f}$.

Now the scaling factors are used to transform $\psi_{\textnormal {c+f}}$ and $\psi_{\textnormal f}$ into their area-normalised versions, $\Psi_{\textnormal {c+f}}$ and $\Psi_{\textnormal f}$, that would represent the frequency functions expected for unit area on the sky (let's say, stars per (mas yr$^{-1}$)$^2$ and per solid angle unit).

\begin{equation}
\label{normalised}
\Psi_{\textnormal {c+f}}=\frac{1}{A_{\textnormal {c+f}}}\psi_{\textnormal {c+f}}
\\
\Psi_{\textnormal {f}}=\frac{1}{A_{\textnormal {f}}}\psi_{\textnormal {f}}
\end{equation}

These new functions are independent of the (arbitrary) areas selected for the regions and can be directly compared. 

The next step will be to deduce the area-normalised frequency function expected for the non-field population. This population, that we will label as 'c' (because most often it will correspond to a cluster), is computed as the simple difference of the two previous ones:

\begin{equation}
\label{clusterff}
\Psi_{\textnormal {c}} = \Psi_{\textnormal {c+f}}-\Psi_{\textnormal {f}} = 
\frac{1}{A_{\textnormal {c+f}}}\psi_{\textnormal {c+f}}-
\frac{1}{A_{\textnormal {f}}}\psi_{\textnormal {f}}
\end{equation}

In a final manipulation of the frequency functions, the scaled versions are deduced. To do this, we first have to define the volumes of the frequency functions as the direct sum of the tabulated values. For the raw functions, we have:

\begin{equation}
\label{volraw}
v_{\textnormal {c+f}}=\sum \psi_{\textnormal {c+f}} \\
v_{\textnormal {f}} =\sum \psi_{\textnormal {f}} 
\end{equation}

The same way, for the area-normalised functions f and c we have:

\begin{equation}
\label{volnorm}
V_{\textnormal {f}}=\sum \Psi_{\textnormal {f}} \\
V_{\textnormal {c}} =\sum \Psi_{\textnormal {c}} 
\end{equation}

The scaled frequency functions are defined in such a way that their volumes are equal to the number of stars of their respective populations:

\begin{equation}
\label{scaledff}
{\overline{\Psi}}_{\textnormal f}=\frac{N_{\textnormal{f}}}{V_{\textnormal{f}}}\Psi_{\textnormal{f}} \\
{\overline{\Psi}}_{\textnormal c}=\frac{N_{\textnormal{c}}}{V_{\textnormal{c}}}\Psi_{\textnormal{c}}
\end{equation}

It can be easily shown that the first of these scaled frequency functions (the function for the field population, f) is equivalent to scaling to sample size $N_{\textnormal{f}}$ the raw function: 

\begin{equation}
{\overline{\Psi}}_{\textnormal f}=
\frac{N_{\textnormal{f}}}{v_{\textnormal{f}}}\psi_{\textnormal{f}} =
\frac{A_{\textnormal{t}} n_{\textnormal{f}}}{A_{\textnormal{f}} v_{\textnormal{f}}}\psi_{\textnormal{f}}
\end{equation}

However, the scaled function for the non-field population, c, would be much more complicated to reduce to directly observed quantities. 

Scaled frequency functions can be regarded as expressed in units of stars per (mas yr$^{-1}$)$^2$, i.e., for each population, they measure the number of stars found at each of the cells of the VPD.

\subsubsection{Noise and membership probabilities}
\label{prob}
The assumptions enumerated in Sec. \ref{assumptions} cannot be perfectly fulfilled. Spatial and kinematical uniformity of the field is only approximate on the real sky, a situation that worsens for larger fields of view. Cluster populations not always stand out clearly as an overdensity on the plane of the sky, mainly for faint clusters, and at the outer areas of even dense clusters. Irregularities in the field spatial distribution may reduce the contrast of the cluster against the field. Finally, open clusters are much more extended than previously thought, with wide coronas that make them not always easy to find areas devoided of cluster members. As a result, the frequency functions predicted by Eqs. \ref{clusterff} and \ref{scaledff} for the non-field population, c, are affected by some noise. This noise level can be estimated from the negative values of the function $\overline{\Psi}_{\textnormal{c}}$, that obviously have no physical meaning. Assuming that these negative values are caused by noise, and that this noise has the same distribution towards positive values, the noise level can be identified with: 

\begin{equation}
\label{noise}
\sigma_{-}=\sqrt{\frac{\sum_{-}\overline{\Psi}_{\textnormal{c}}}{M}}
\end{equation}

Where the summation is performed over the cells with negative value, and $M$ is the number of those negative-valued cells. 
This estimation allows to evaluate the signal-to-noise ratio at each cell of this function.

Finally, membership probabilities can be assigned to each cell of the VPD over the working space, as follows:

\begin{equation}
\label{probabilities}
P=\frac{\overline{\Psi}_{\textnormal {c}}}{\overline{\Psi}_{\textnormal{c}}+\overline{\Psi}_{\textnormal{f}}}
\end{equation}

Normally, the probability function $P$ will be set to zero at those cells of the VPD where the signal to noise ratio of function $\overline{\Psi}_{\textnormal{c}}$ is below some threshold $\gamma \sigma_{-}$, where $\gamma$ is a parameter that has to be adjusted in each case, depending on the characteristics of the sample.  

The simplicity of Eq. \ref{probabilities} is more apparent than real, in the sense that it is not straightforward to express that definition in terms of the observables and the raw frequency functions.

Given any star from the original sample, the probability for it belonging to the non-field population is identified with the value of function $P$ at the cell of the VPD occupied by the star, according to its proper motions. The expected number of non-field stars, $N_{\textnormal{c}}$ (Eq. \ref{expectedmembers}) may serve as a guide to set a probability cut-off. 

\section{The \texttt{Clusterix} workflow}
\label{workflow}
The general, theoretical frame formulated in Sec. \ref{nonparametric}, is implemented into \texttt{Clusterix} 2.0 in the way described as follows.

\subsection{Data query}
The first step of the procedure is the definition of the data source and the working space.

Users can gather information on a particular region in the sky from the following catalogues: {\em Gaia} DR2 \citep{Brown2018}, TGAS \citep{Prusti2016}, PPMXL \citep{Roeser2010}, UCAC4 \citep{Zacharias2012}, UCAC5 \citep{Zacharias2017} and HSOY \citep{Altmann2017}. 
Queries can be made using equatorial coordinates and a search radius. Any object identifier that can be resolved by the Sesame\footnote{http://cds.u-strasbg.fr/cgi-bin/Sesame} service can also be used. In both cases, the query can be filtered specifying a range of magnitudes ($G$, $G_{BP}$ or $G_{RP}$ in the case of {\em Gaia}, $J$,$H$ or $K_S$ for 2MASS and so on). 
In the case of {\em Gaia} DR2 users can select as well the Q-Filter that follows the equations C.1 and C.2 from \citet{Lindegren2018}. These filters help to work with an astrometrically clean subset (see also \citealt{arenou2018}).

\begin{figure*}
\includegraphics[width=\textwidth]{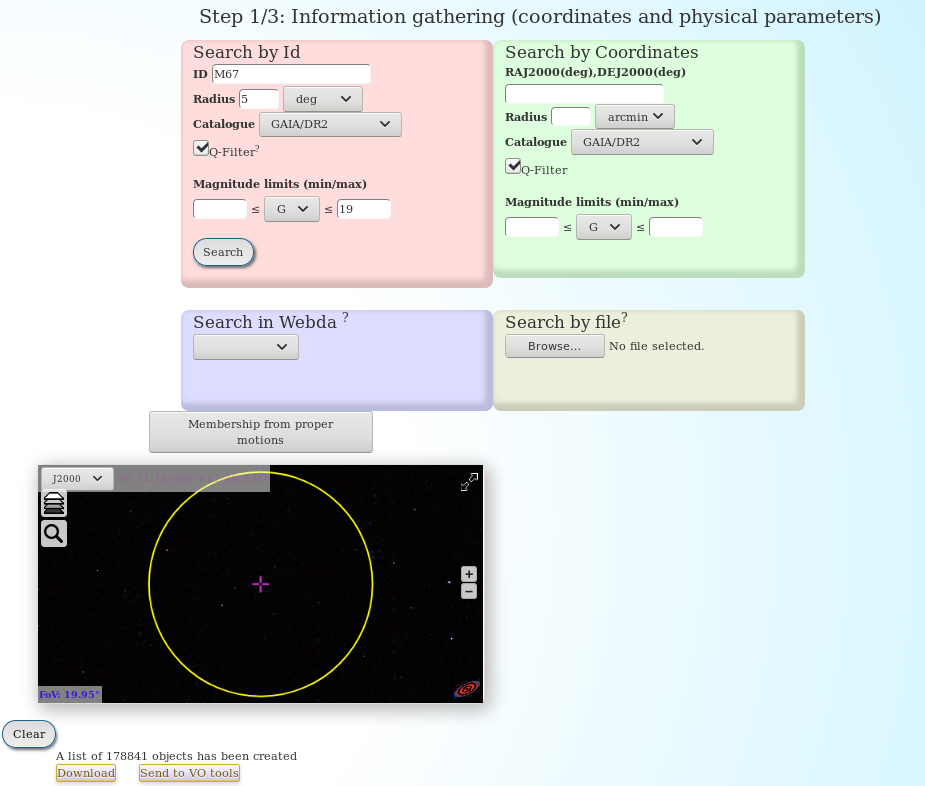}
\caption{Step 1 of \texttt{Clusterix} 2.0. for the case of M~67. The catalogue, the radius and the magnitude limit are chosen in the upper panels. Results can be downloaded or sent to Virtual Observatory tools. When the number of stars is bigger than 40\,000, only a yellow circle is drawn to enclose the searched region. See text for more details. See Sect.~4.1 for details on the parameters selection for this cluster.} 
\label{fig:step 1}
\end{figure*}

Alternatively, users can upload ASCII files with their own data. This comes very handy for custom datasets or synthetic data. Files must have comma separated values (CSV files) and the fields can be in any order. Mandatory fields and labels are: ID, RAJ2000, DEJ2000, pmRAcosdelta and pmDEC.
If information on magnitude (MAG) and proper motion uncertainties (epmRAcosdelta, epmDEC) is available in the input file, it will be possible to use them in the subsequent steps. The magnitude is used only to apply limits to the workplace if desired. Proper motion errors are also only used to apply limits that allow to discard data of bad quality, if present, but do not play any additional role in the definition of the frequency functions themselves. 
Extra columns present in the input file will be accepted by Clusterix 2.0 but ignored for calculations. By hovering the mouse over the question mark next to this section in the web (Fig.~\ref{fig:step 1}), the user can download some examples of input files that can be used as template. The privacy of user data is guaranteed, since the files uploaded by the users are not kept in the system.

Finally, the main source of data of the previous version, \texttt{Clusterix} 1.0, the WEBDA database\footnote{http://webda.physics.muni.cz/}, can still be used as a source of information on positions and proper motions.

The result of the query is visualized using Aladin Lite\footnote{http://aladin.u-strasbg.fr/AladinLite/doc/}, where the objects obtained from the query are plotted as red diamonds. The user can zoom and pan, as well as select different background images. As the representation of tens of thousand objects is quite demanding from a computational point of view, we limit the individual representation of objects (the red diamonds) to $40\,000$. If the result of a query exceeds this amount, then just a yellow circle enclosing the search region is drawn.

Then, the result of the query can be sent to the next step ("Membership from proper motions"), or be downloaded as a CSV file or sent to VO applications using the Simple Application Messaging Protocol (SAMP\footnote{http://www.ivoa.net/documents/REC/App/SAMP-20090421.html}) for further analysis.

\subsection{Selection of the {\it cluster+field} and {\it field} regions }

The second step is to select the {\em cluster+field} (c+f) and {\em only field} (f) regions mentioned in Sec.~\ref{assumptions}. The f and c+f regions may be, in principle, arbitrary in shape. In fact, they do not even need to be connected spaces (each of them may be made up from separate pieces), nor simply connected spaces (they even may have holes). The definition of these areas is one of the most critical decisions to take by the user, and \texttt{Clusterix} 2.0 offers several ways to interactively shape and reshape these areas. 

\begin{figure*}
\includegraphics[width=\textwidth]{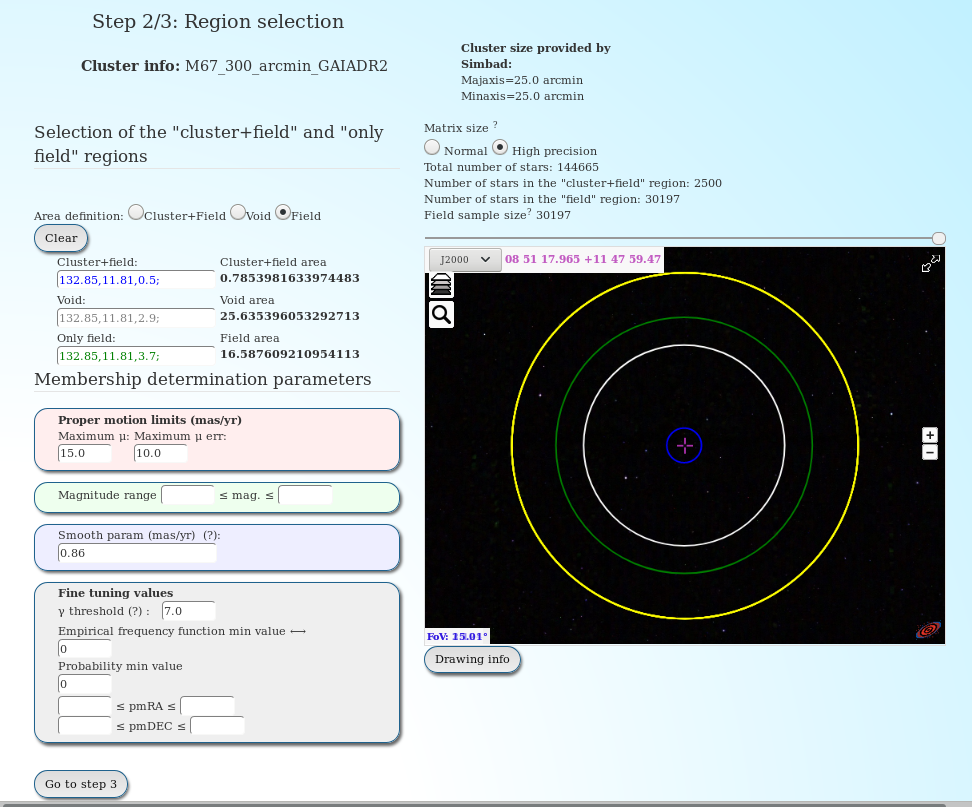}
\caption{Step~2 where all the fine tuning parameters of the \texttt{Clusterix} procedure are set for M~67. In this step c+f (blue), f (green), void (white), and total (yellow) areas are defined and parameters are chosen interactively. Tests can be done in normal precision while high precision is used here for final results. See text for more details.}
\label{fig:step 2}
\end{figure*}

The simplest option relies on mouse clicks to draw circles that define the c+f (inside the blue circle in Fig.~\ref{fig:step 2}) and f (inside the green circle in the same figure down to the following circle) regions. The system also includes an easy way to set up a "buffer" void area around the c+f region that will be excluded from both areas (the area marked between the white and the blue circles in the figure). This is to define a clean field area avoiding an intermediate region that could still have a significant number of cluster members, so no stars inside that buffer are taken into account for the frequency functions calculations. See an example of the three areas in Fig.~\ref{fig:step 2}.

Alternatively, the user can specify the circular areas directly writing their parameters (in decimal degrees) in the corresponding boxes (format: "ra,dec,radius;").

These circles do not need to be concentric and other distributions are possible (see i.e. Sect.~\ref{Rup26}). If there is an overlap area of the two circles (f and c+f), that area is assigned only to the c+f and excluded from the f sample. The buffer (void area) is as well excluded from the f sample.  

Once the areas are chosen, the star sample can be further filtered in this second step according to the "Membership determination parameters": 
\begin{enumerate}
\item Proper motion limits. Maximum value of the total proper motion (to reduce the size of the VPD for the working space, discarding since the beginning objects that clearly cannot belong to the expected cluster population) and maximum value of the total proper motion error (to remove data of dubious quality). 
\item Magnitude range to further limit the selection (with respect the magnitude chosen in Step~1), if desired.
\item Smoothing parameter. \texttt{Clusterix} 2.0
proposes a default value for the smoothing parameter $h$ on the basis of Eq.~\ref{silverman} applied to the c+f sample, and this can be modified by the user. The same $h$ value is used to compute both frequency functions. 
As $h$ represents the radius of the kernel window (i.e., the sigma of the Gaussian kernel $K$), a large value would blur out the details of the frequency functions, while a small value would yield noisy results, because a very low number of individuals would receive significant weights. 
\item Fine tuning values. Here we mainly fix the value of $\gamma$ explained in previous section. To avoid meaningless probability values, \texttt{Clusterix} 2.0 restricts the probability calculations to stars with densities $\gamma$ times above the noise.  
\end{enumerate}

All these parameters, shown in Fig.~\ref{fig:step 2}, can be interactively chosen depending on the data used and/or the region studied, until satisfactory results are obtained.

\subsection{Determination of the empirical frequency functions}

The empirical determination of the frequency functions is performed according to the formalism of Sec.~\ref{nonparametric}. The assumptions stated in Sec.~\ref{assumptions} are critical for the membership determination. Assumption (i) is usually fulfilled, but may be problematic for very large work spaces. Finding regions that fulfill assumption (iii) may be more difficult because, a priori, there is no information about the true extension of the cluster, and even its location on the sky may be uncertain. It is, thus, important to perform tests with areas of very different sizes searching for a reasonable trade-off between cleanness (absence of a significant amount of cluster members) and signal-to-noise ratio (working area not too small, to avoid small number statistics). 

\begin{figure*}
\includegraphics[width=\textwidth]{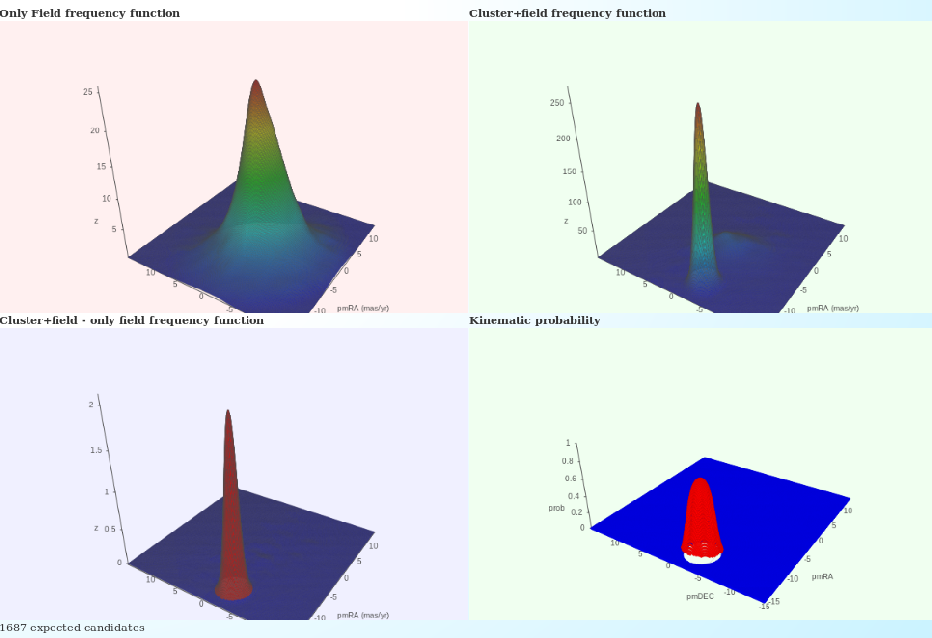}
\caption{Interactive results of Step~2 of the \texttt{Clusterix} procedure for M~67. Visualization of the frequency functions helps to decide on the best set of areas and parameters: f and c+f (top) and c and probability (bottom). The user can rotate and change the angle of the plots as well as click in any cell to know its values. See text for more details.}
\label{fig:step 2b}
\end{figure*}

The frequency functions are empirically determined as tables that contain cells defined in the VPD. To facilitate the first, tentative tests, \texttt{Clusterix} 2.0 works with two different matrix sizes to discretize the VPD: 100 $\times$ 100 ("Normal") or 300 $\times$ 300 cells ("High precision"). The first size is useful for fast queries, while the second one allows a more precise data representation and results. See Figs.~\ref{fig:step 2} and ~\ref{fig:step 2b} for the example in High precision.

One of the interesting capabilities of \texttt{Clusterix} 2.0 is that, after selecting the c+f and f regions, the tool provides a visualization of the resulting empirical, raw frequency functions $\psi_{\textnormal{c+f}}$ and $\psi_{\textnormal{f}}$. 
This allows the user to evaluate the suitability of the selected regions. Usually, if a cluster is present, $\psi_{\textnormal{f}}$ should display a broad shape with only one maximum, and $\psi_{\textnormal{c+f}}$ would include that same profile with an additional, normally sharper, peak due to the cluster (Fig. \ref{fig:step 2b}, top panel). Figures can be interactively rotated and zoomed in and out. Also, by placing the cursor on any point of the graphics, its coordinates (pmRA, pmDEC, frequency value) are displayed. 

There is another parameter that can be fixed for the case of big sample size: "Field sample size". Frequency function calculations can be optimized by reducing the number of stars used. Statistically we only need a subset of stars to obtain the relevant information that is needed for determining the functions. Lower values of this parameter can also be used for quick testing. 
The suggested upper limit at $50\,000$ stars is set as no significant differences are obtained with greater numbers and the calculations can then take too long to complete. See example in Sect.~4.2.

\subsection{Visualizing the results}

After each redefinition of the c+f and f regions, the system recomputes and displays the empirical frequency functions and, also, calculates and shows the scaled cluster frequency function $\overline{\Psi}_{\textnormal{c}}$ (Eq. \ref{clusterff}) and probability distribution $P$ (Eq. \ref{probabilities}). In these two graphs, high signal to noise ratio areas are marked in red (see Fig. \ref{fig:step 2b}, bottom panel). According to what was stated in Sec. \ref{prob}, $P$ is null for those cells of the VPD with $\overline{\Psi}_{\textnormal{c}} < \gamma \sigma_{-}$, where the $\gamma$ factor is set by the user.

\begin{figure*}
\includegraphics[width=\textwidth]{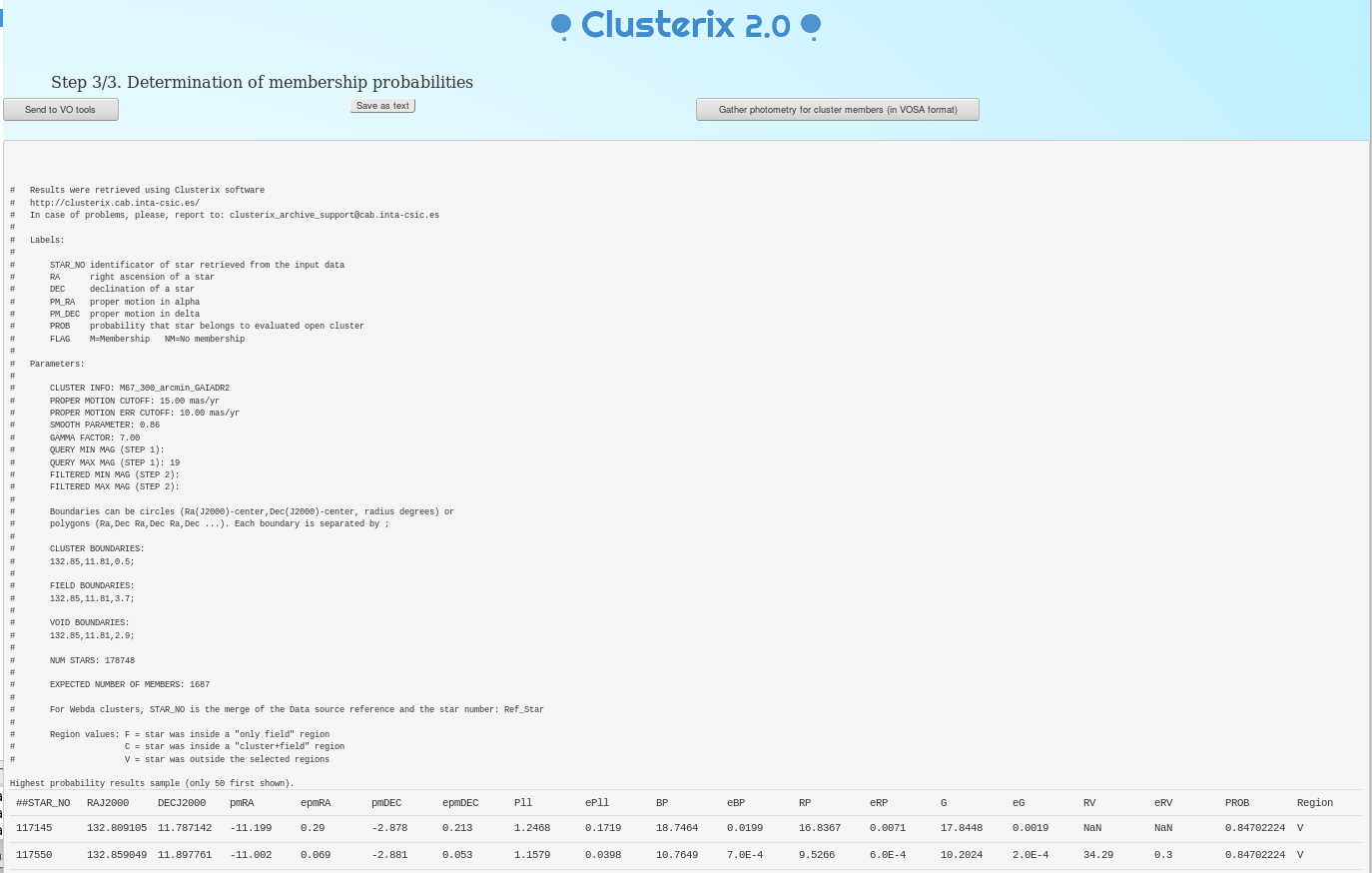}
\caption{Step~3 of the \texttt{Clusterix} procedure for M~67. All the parameters are saved together with the results in the file downloads as text while the star list can be directly sent to Topcat/Aladin via SAMP or a search of values from VOSA can be performed. Only the first row of the results are shown. }
\label{fig:step 3}
\end{figure*}

Once the configuration of c+f and f regions and the other parameters are considered satisfactory enough, the user executes the command "Go to step 3" to trigger the computation of the results for the whole work space and the catalogue is created. The results are displayed in a new page and they can be downloaded as a CSV file, or sent to any VO tool (e.g. Topcat) using the SAMP protocol. A summary of the results is shown directly on the browser along with the parameters used for the calculations, so the experiment can be replicated again in the future if needed (see Fig.~\ref{fig:step 3}).

An estimation of the number of cluster members is provided, based on the expected overpopulation of the cluster region provided in step~2, but this number may not be exact due to a inexact region determination or a non homogeneous field distribution. This  expected number of members is given as a result $N_{\textnormal{c}}$ and it would correspond to the first $N_{\textnormal{c}}$ stars with higher probability of being members. But the full sample is included in the given file with their individual probabilities, allowing the users to apply their own decision using further criteria.

\begin{figure*}
\includegraphics[width=\textwidth]{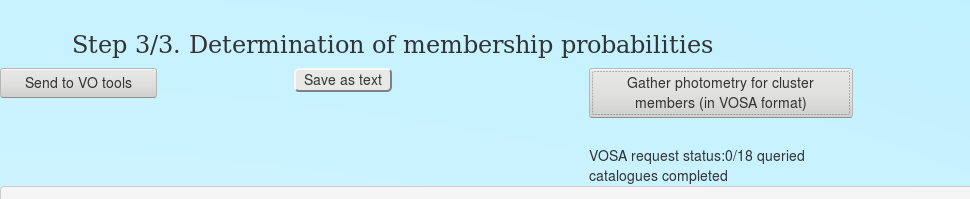}
\includegraphics[width=\textwidth]{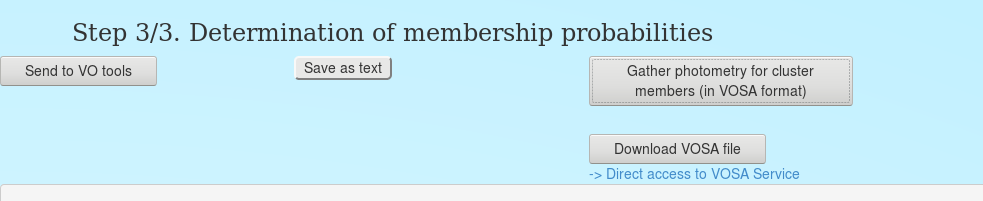}
\caption{\texttt{Clusterix} 2.0. Step 3 for the gathering of photometry with VOSA. Top panel when the system begins the search. Bottom panel, once the search is finished the user can download the photometry in the correct format for VOSA uploading.}
\label{fig:step 3b}
\end{figure*}

For those $N_{\textnormal{c}}$ stars with higher probability of being cluster members, it is possible to gather additional photometric data using the VOSA\footnote{http://svo2.cab.inta-csic.es/theory/vosa/} service. The user can download this photometry, make their own selection, and upload it again to work with VOSA. \texttt{Clusterix} 2.0 implements a simple login system in order to access to past VOSA queries. Also, as they can take a long time, the system periodically checks if a photometry query is finished and sends an email to the user. This user authentication system is not needed for other \texttt{Clusterix} 2.0 functionality. See Fig.~\ref{fig:step 3b}.

\section{The science cases}
\label{science}

The tool discussed in the previous sections is applied to a few specific cases using {\em Gaia} DR2 to demonstrate its performance in easy and complex scenarios. We have chosen five areas of clusters of different sizes, ages and distances, as well as areas where more than one cluster is present: from known overlapping clusters to unexpected new ones.

We follow the recommendations by \citet{arenou2018} and \citet{Lindegren2018}, Eqs. C.1 and C.2, for an astrometrically clean subset. Although the {\em Gaia} limiting magnitude is $G =$ 21.0, the median uncertainty for the bright sources ($G <$ 15 mag) is 0.06 mas yr$^{-1}$ in proper motions, and goes up to 0.3 mas yr$^{-1}$ for $G =$ 18. 

Depending on the cases, we only consider sources brighter than $G =$ 18 or 19. Fainter objects have larger uncertainties that can blur the separation between cluster and field in the proper motion parameter space. Also, this cut in magnitude significantly reduces the amount of sources to be managed as 80\% of the {\em Gaia} DR2 sources are fainter than G $\sim$18. \

The positions and main parameters of the stars in the areas of the clusters discussed in this section are given in Table~\ref{tab:generalresults} in electronic format only.

\subsection{NGC~2682 (M~67)}
\label{M67}

\begin{figure}
	\includegraphics[width=\columnwidth]{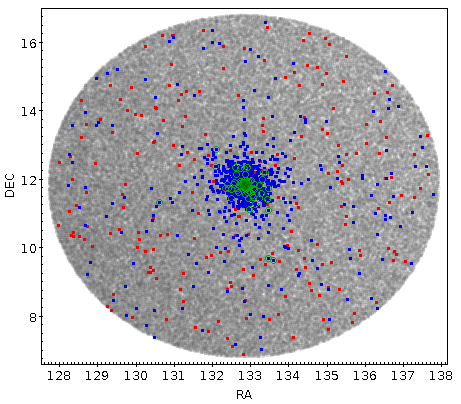}
    \caption{An area of 5\,$\deg$ radius around M~67 is studied. In gray all stars up to magnitude $G$= 19 with proper motions from {\em Gaia} (178\,748). The selection made by \texttt{Clusterix}: 1\,699 stars (in red) is refined with a cut in parallax: 1\,440 members (in blue). Superimposed in green the 84 stars with compatible {\em Gaia} radial velocity.}
    \label{fig:M67_area}
\end{figure}  
\begin{figure}
	\includegraphics[width=\columnwidth]{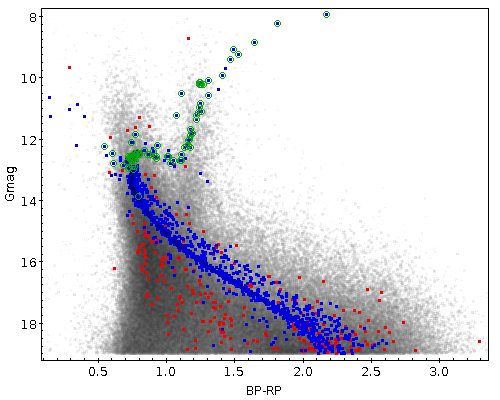}
    \caption{Colour-Magnitude diagram of M~67. In gray all stars in a 5\,$\deg$ radius around the centre up to magnitude $G$= 19 with proper motions (178\,748). The selection made by \texttt{Clusterix}: 1\,699 stars (in red and blue) is refined with a cut in parallax: 1\,440 members (in blue). Superimposed in green circles the 84 stars with compatible {\em Gaia} radial velocity.} 
    \label{fig:HR_M67}
\end{figure}

M~67 is an old cluster \citep[$\sim$3.6 Gyr][]{Bossini2019} at about 900 pc with a near solar metallicity and low reddening. M~67 is one of the best studied open clusters, considered a cornerstone of stellar astrophysics and used as a calibrator in many surveys. However there was no study covering a large area in spite of some studies on its corona showing that it is an extended cluster. \cite{Babusiaux2018} studied an area of 1 degree with $G<$20 and found 1\,520 members. 

We study the data from \textit{Gaia}~DR2 in a radius of 5 degrees from the cluster centre and magnitude limit ($G<$19) (178\,748 stars). The extended halo of this cluster has been studied by \cite{Carrera2019b}, based on the data explained here. In the search for the farthest objects, their study extended up to 10 degrees (but $G<$ 18) and shows that the cluster radial distribution $r_{85\%}$ = 38\arcmin, well under our chosen 5 degrees. The bigger the field, the worse representation of the field (see Appendix A on their paper) and the noisier the solution. See the area covered in this study in Fig.~\ref{fig:M67_area}. 

In step~2 of \texttt{Clusterix}, we apply a constraint in proper motions ($\mu \leq$ 15\,mas yr$^{-1}$)
and in proper motions errors (below 10\,mas yr$^{-1}$) to avoid outliers and limit the computation time (see Fig.~\ref{fig:step 2}). 

\texttt{Clusterix} 2.0. gives us an estimated number of members of 1\,687, what corresponds to a probability cut at 0.79479. With this probability cut, we get 1\,699 candidate members using only proper motions. If we apply a $1~\sigma$ cut in parallax 
to that selection we obtain a final list of 1\,440 member stars (see Fig.~\ref{fig:HR_M67}). From this sample we check those with radial velocities from {\em Gaia} DR2 (88 stars) and use it only to calculate a median radial velocity with a sigma clipping, resulting in $v_{\rm rad} = 34.5\pm2.0$\,km s$^{-1}$ from 84 stars.
\cite{soubiran2018} find a $v_{\rm rad} = 33.80\pm1.06$\,km s$^{-1}$ from 64 member stars. 

The average proper motion and parallax of the cluster are:
$\mu_{\alpha*} = -10.99\pm0.28$\,mas yr$^{-1}$; $\mu_{\delta} = -2.94\pm0.26$\,mas yr$^{-1}$; $\varpi =$ 1.13$\pm$0.11\,mas from 1\,440 member stars. The average parallax from the 84 radial velocity members is $\varpi =$ 1.16$\pm$0.06\,mas.

All the errors are the standard deviation of the distribution. 
Comparing this result with the 691 member stars found in the study on the 30{\arcmin} central area with a magnitude limit of $G=18$ from \citet{tristan2018b}, we find 677 members in common (98\% agreement). From the 1\,251 members by \cite{Babusiaux2018} with $G<$19, we have 1\,116 members in common (89\% agreement). The comparison of the final values are in Table~\ref{tab:m67}.

\begin{table*}
 \caption{Comparison of our results for M~67 with other authors.}
 \label{tab:m67}
 \begin{tabular}{llllll}
  \hline
   & $\mu_{\alpha*}$ & $\mu_{\delta}$ & $\varpi$ 
   & $N_{\textnormal{mem}}$ & $G_{\rm lim}$\\
      & (mas yr$^{-1}$) & (mas yr$^{-1}$) & (mas) &  & (mag) \\
  \hline
  This study & $-$10.99$\pm$0.28 & $-$2.94$\pm$0.26 
  &  1.13$\pm$0.11 & 1440 & 19\\
  \citet{tristan2018b} & $-10.986\pm$0.193 & $-2.964\pm$0.201  & 1.135$\pm$0.051 & 691 & 18 \\
  \cite{Babusiaux2018} & $-$10.974$\pm$0.006* & $-$2.940$\pm$0.006*  & 1.133$\pm$0.001* & 1520 & 20 \\
  \hline
 \end{tabular}
 
  * in the case of \cite{Babusiaux2018} the errors quoted are uncertainties and not standard deviations.

\end{table*}  

\subsection{NGC~2516}
\label{N2516}

\begin{figure}
	\includegraphics[width=\columnwidth]{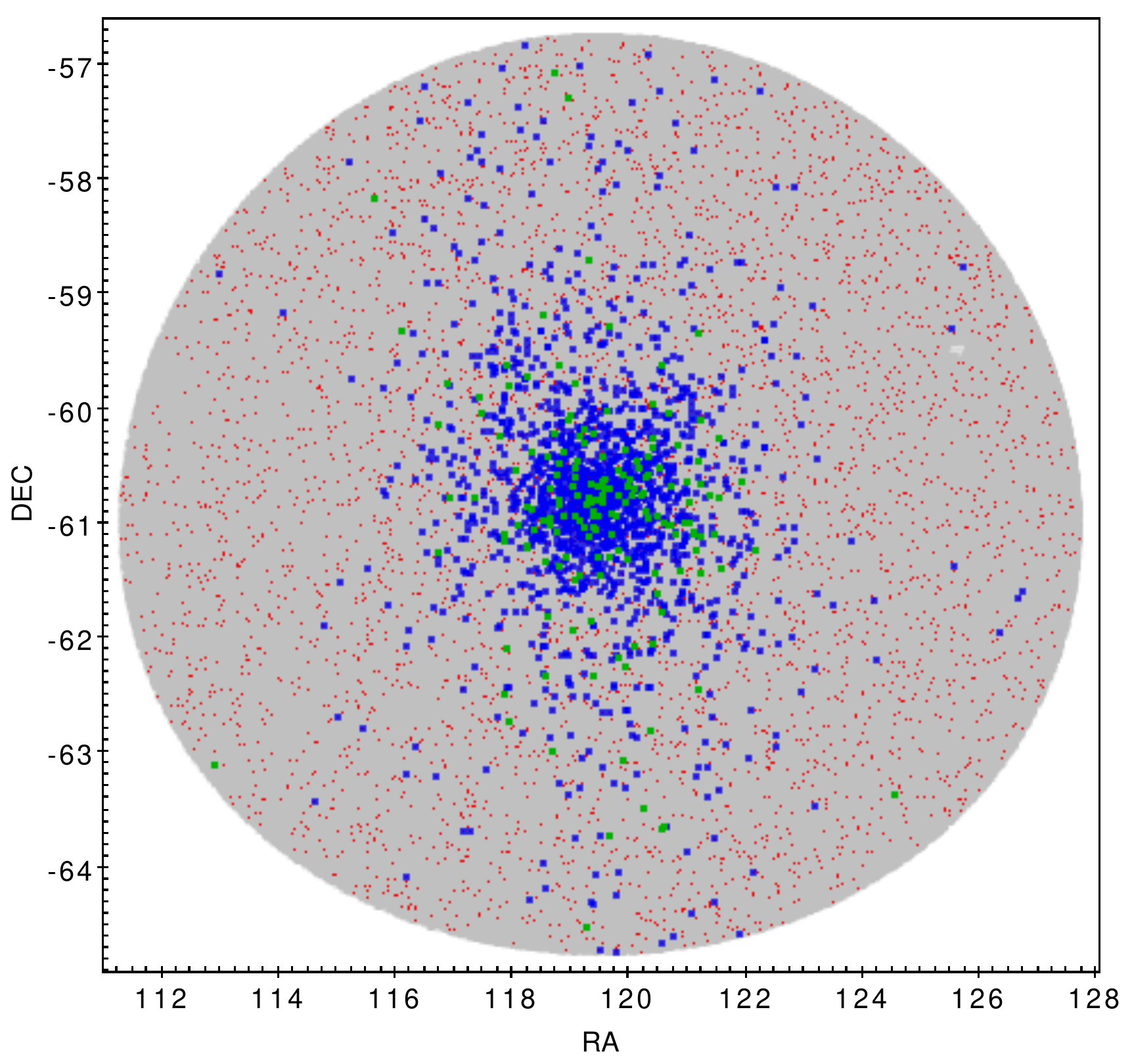}
    \caption{Diagram of NGC~2516 area. In translucent gray all stars in a 4 degree radius around the centre (552\,583 stars). The selection made by \texttt{Clusterix}: 6\,210 stars (in red and blue), refined with parallax: 1\,819 members (in blue), where 175 in green show compatible {\em Gaia} radial velocities. A clear elongated structure is visible.}
    \label{fig:NGC2516_area}
\end{figure}

\begin{figure}
	\includegraphics[width=\columnwidth]{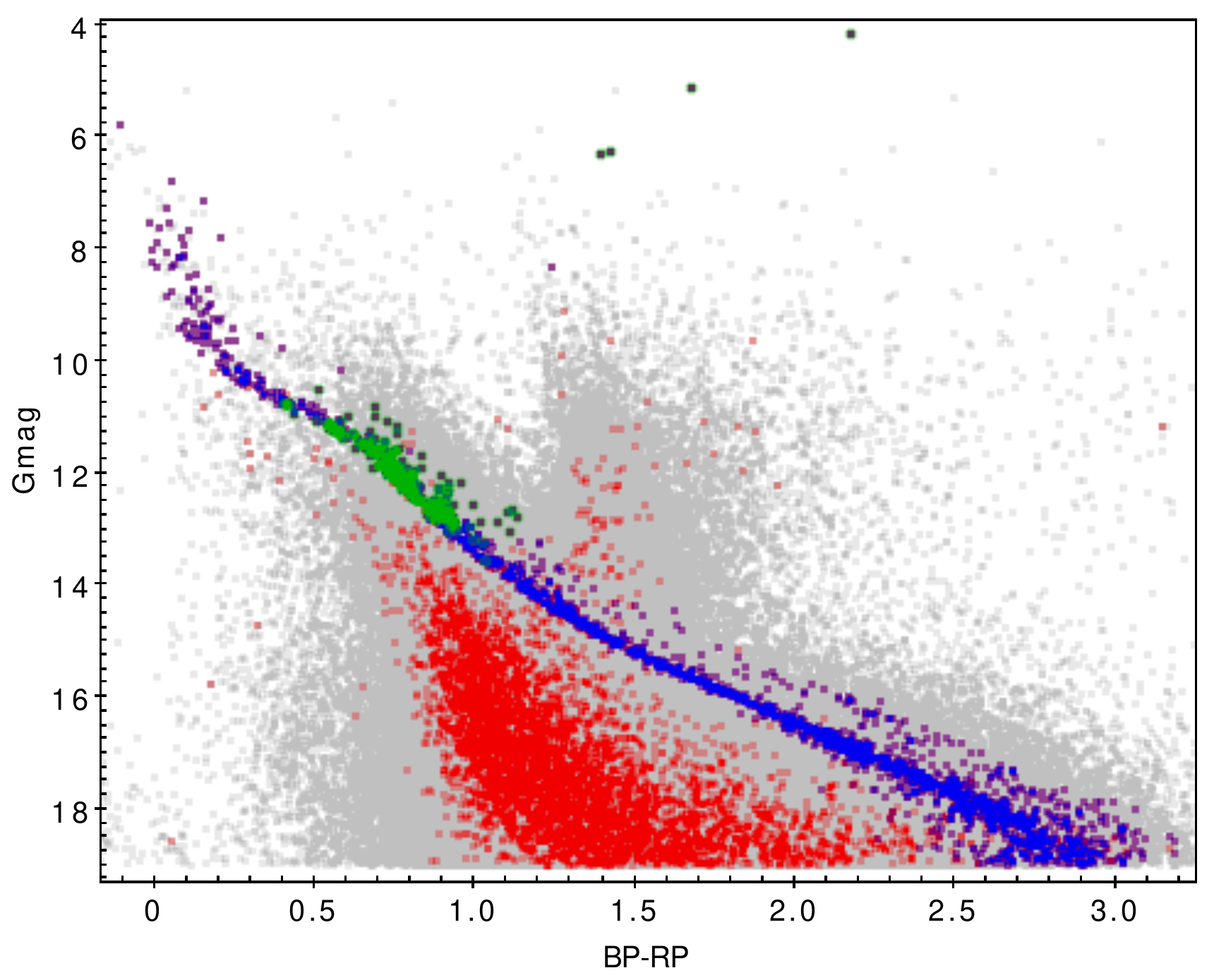}
    \caption{Colour-Magnitude diagram of NGC~2516. In translucent gray all stars in a 4 degree radius around the centre (552\,583 stars). The selection made by \texttt{Clusterix}: 6\,210 stars (in red and blue), refined with parallax: 1\,819 members (in blue), where 175 in green show compatible {\em Gaia} radial velocities.}
    \label{fig:HR_NGC2516}
\end{figure}

\begin{table*}
 \caption{Comparison of our results for NGC~2516 with other authors.}
 \label{tab:N2516}
 \begin{tabular}{llllll}
  \hline
   & $\mu_{\alpha*}$ & $\mu_{\delta}$ & $\varpi$ 
   & $N_{\textnormal{mem}}$ & $G_{\rm lim}$ \\
      & (mas yr$^{-1}$) & (mas yr$^{-1}$) & (mas) & & (mag) \\
  \hline
  This study & $-$4.63$\pm$0.43 & 11.15$\pm$0.35 
  &  2.4$\pm$0.1 & 1\,819 & 19 \\
  \citet{tristan2018b} & $-$4.748$\pm$0.441 & 11.221$\pm$0.345  & 2.417$\pm$0.045 & 798 & 18 \\
  \cite{Babusiaux2018} & $-$4.6579$\pm$0.0075* & 11.1517$\pm$0.0075*  & 2.4118$\pm$0.0006* & 2518 & 20 \\
  \hline
 \end{tabular}
 
  * in the case of \cite{Babusiaux2018} the errors quoted are uncertainties and not standard deviations.

\end{table*}  

NGC~2516 is a young cluster ($\sim$~150 Myr), at a distance of about 400 pc and a $E(B-V)$ $\sim$0.12 mag. It has a known radial structure already visible in Fig D.24 in \citet{Floor2017} from {\em Gaia} DR1 \citep{Brown2016}.

We have studied an area of 4\,$\deg$ radius around the centre and magnitude limit ($G<$19) (552\,583 stars). It is a very populated area and we tested different areas to maximize the contrast between the field and the cluster. The final result (see Appendix for more information on the areas and fine tuning parameters chosen) gives a number of expected members of 2\,537. 
However, in this case, being a very close cluster, the parallax is a greatly discriminant parameter. So we accept all stars with probability greater than 0 (that means Prob > 0.6) and apply a cut in the clear overdensity parallax as it is quite straightforward (2$<\varpi<$3\,mas), 
resulting in 1\,819 member stars. The resulting CM diagram shows the clean selection (see Fig.~\ref{fig:HR_NGC2516}). 

From the 175 stars with compatible {\em Gaia} radial velocities, the average radial velocity of the cluster $v_{\textnormal{rad}} = 24.4\pm2.6$\,km s$^{-1}$.  

The proper motion and parallax of the cluster are:
$\mu_{\alpha*} = -4.63\pm$0.43\,mas yr$^{-1}$; $\mu_{\delta} =$ 11.14$\pm$0.35\,mas yr$^{-1}$; $\varpi =$ 2.4$\pm$0.1\,mas for the 1\,819 member stars. The average parallax from the 175 radial velocity members is $\varpi =$ 2.42$\pm$0.08\,mas.

The member stars are distributed as seen in Fig.~\ref{fig:NGC2516_area}. The elongated distribution is clear even among the 175 radial velocity members. 

Comparing this result with the 798  member stars found in the study on the 1\,$\deg$ central area with a magnitude limit of $G=18$ by \citet{tristan2018b}, we find 696 in common with our member list (87$\%$ agreement). 
From the 2\,125 members by \cite{Babusiaux2018} with $G<$19 in a 2\,$\deg$ area, we have 1\,620 members in common with our final selection (76\%). 
The comparison of the final values are in Table~\ref{tab:N2516}.

Comparing our membership with the radial velocity study by \citet{Bailey2018}, from high resolution spectroscopy of 126 Main Sequence stars in the core 30{\arcmin} of NGC~2516, we have 43 stars in common and a 93$\%$ agreement in membership classification. They give a $v_{\textnormal{rad}} = 24.50\pm0.12$\,km s$^{-1}$ from 81 member stars. 

\subsection{NGC~1817 and Juchert~23}  
\label{N1817}

\begin{figure}
	\includegraphics[width=\columnwidth]{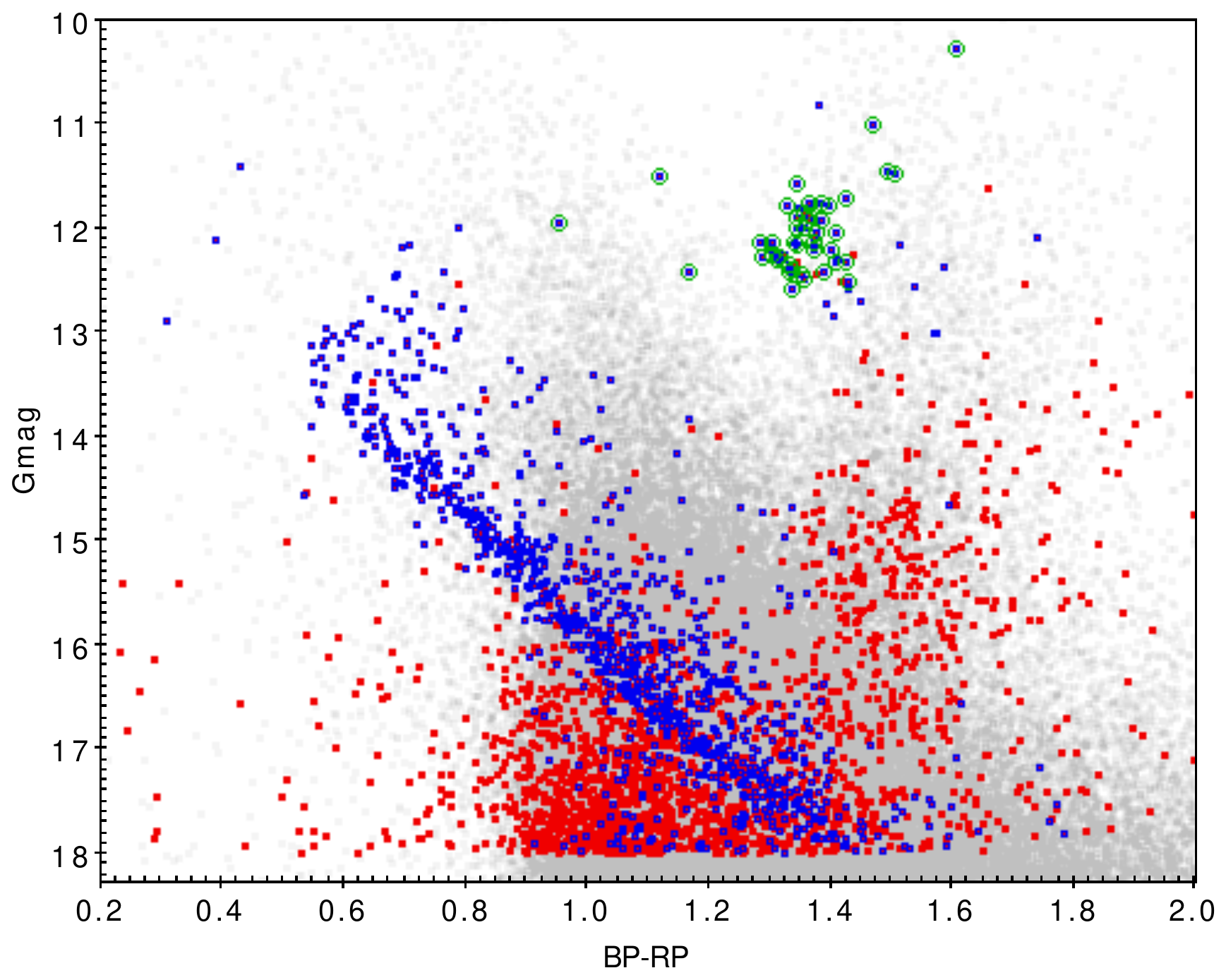}
    \caption{Colour Magnitude diagram of NGC~1817. In translucent gray all stars in a 4 degree radius around the centre (102\,313 stars). The selection made by \texttt{Clusterix}: 2898 stars (in red and blue), refined with a cut in parallax: 1000 members (in blue), where 43 in green show compatible {\em Gaia} radial velocities. }
    \label{fig:HR_N1817}
\end{figure}

In spite of the work by \cite{BalaguerNunez2004, BalaguerNunez2004b} about the open cluster region NGC~1817 where only one very extended cluster was found and the asterism NGC~1807 was confirmed as a non physical cluster, subsequent general catalogues have included NGC~1807 \citep{Kharchenko2016, KroneMartins2010}, and even giving physical parameters (age, extinction...). The precision and completeness of the {\em Gaia} data drives us to study again this area. 

We have studied an area of 2\,$\deg$ radius around NGC~1817 with $G<$18 (102\,313 stars) to check if any sign of NGC~1807 is present. This cluster is far, located at around 1700~pc, and because of its small proper motion it is more difficult to disentangle from the field than the previous examples of NGC\,2682 and NGC\,2516. 
\texttt{Clusterix} finds 2\,898 members in the area. For a cleaner list of members, we cut around the overdensity in parallax 
(0.4$<\varpi<$0.7\,mas)
what makes the cluster very conspicuous in the Colour Magnitude diagram. From the 1000 members in that selection (see Fig.~\ref{fig:HR_N1817}), there are 43 stars with coherent {\em Gaia} radial velocities, giving an average of $v_{\rm rad} = 66.3\pm$1.9\,km s$^{-1}$.   

The proper motion and parallax of the cluster are:
$\mu_{\alpha*}=$ 0.48$\pm$0.19\,mas yr$^{-1}$; $\mu_{\delta}=$ $-$0.89$\pm$0.17\,mas yr$^{-1}$; $\varpi=$ 0.54$\pm$0.07\,mas for 1000 member stars. The average parallax from the 43 radial velocity members is $\varpi=$ 0.54$\pm$0.04\,mas.

Comparing this result with the  member stars found in the study on the 1\,$\deg$ central area with a magnitude limit of $G=18$ by \citet{tristan2018b}, we find 447 member stars in common (97$\%$ agreement). From the 169 astrometric members found by \cite{BalaguerNunez2004} with $V<$14.5 in a 1.5\,$\deg$x1.5\,$\deg$ area, we have 92 members in common (54$\%$ of agreement), what shows the superior quality of the {\em Gaia} data. The comparison of the final values are in Table~\ref{tab:N1817}.

\begin{table*}
 \caption{Comparison of our results for NGC~1817 with other authors.}
 \label{tab:N1817}
 \begin{tabular}{llllll}
  \hline
   & $\mu_{\alpha*}$ & $\mu_{\delta}$ & $\varpi$ 
   & $N_{\textnormal{mem}}$ & $G_{\rm lim}$ \\
      & (mas yr$^{-1}$) & (mas yr$^{-1}$) & (mas) &  & (mag)\\
  \hline
  This study & 0.48$\pm$0.19 & $-$0.89$\pm$0.17 
  &  0.54$\pm$0.07 & 1000 & 18 \\
  \citet{tristan2018b} & 0.485$\pm$0.118 & $-$0.89$\pm$0.1  & 0.551$\pm$0.056 & 460 & 18 \\ 
  \cite{BalaguerNunez2004} & 0.29$\pm$0.10 & $-$0.96$\pm$0.07  & - & 169 & 14.5 \\
  \hline
 \end{tabular}
\end{table*}  
 
\begin{figure}
	\includegraphics[width=\columnwidth]{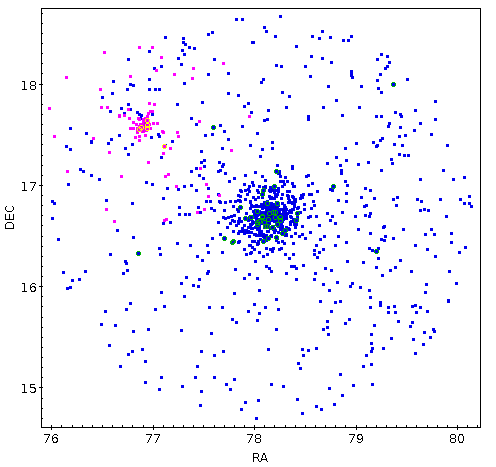}
    \caption{Area of NGC~1817 (members in blue) and Juchert~23 (members in pink). Radial velocity members marked as open circles (green for NGC~1817, yellow for Juchert~23.)}
    \label{fig:N1817nJuc23_area}
\end{figure}

\begin{figure}
	\includegraphics[width=\columnwidth]{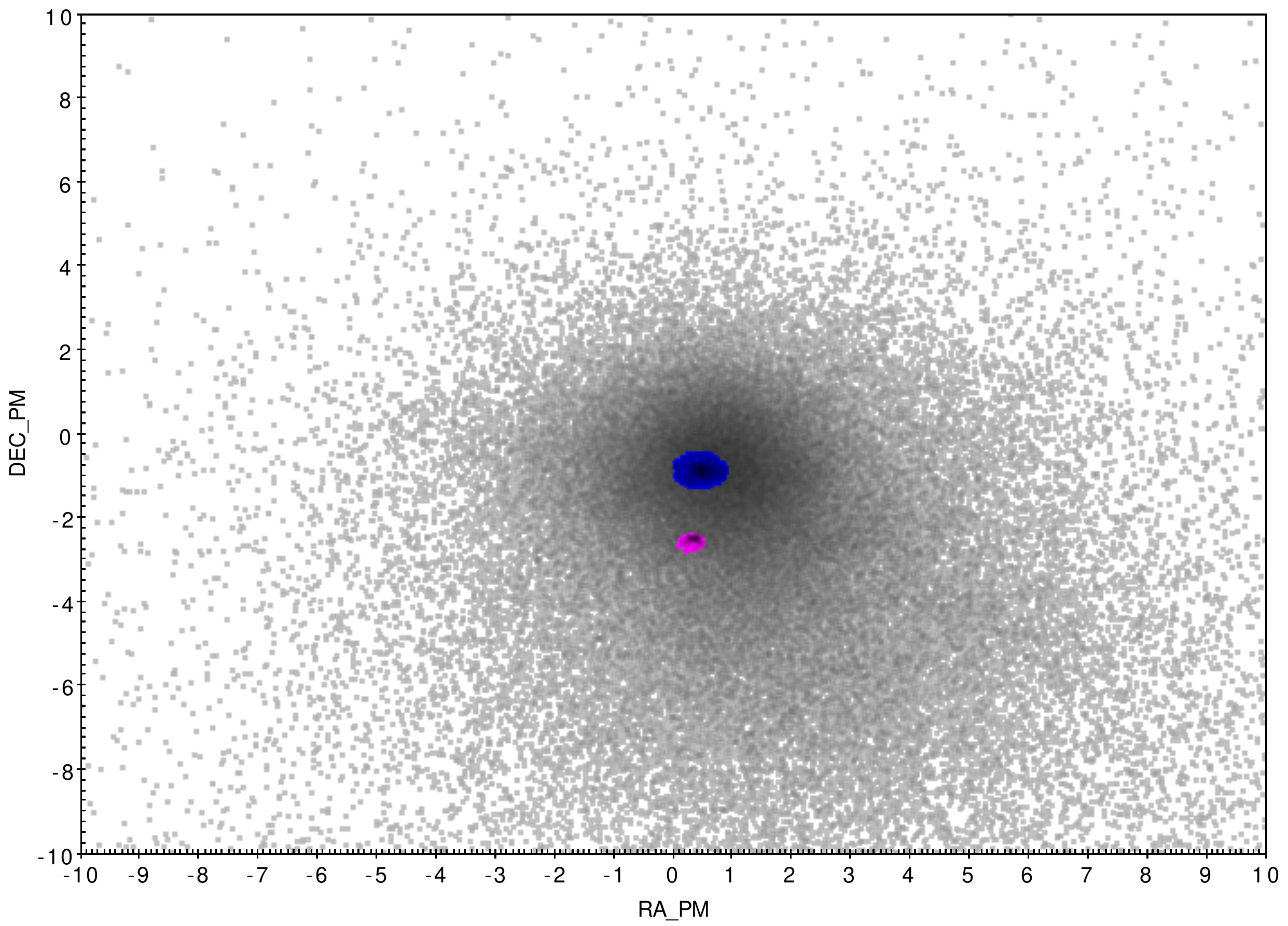}
    \caption{Vector point diagram of NGC~1817 (members in blue) and Juchert~23 (members in pink).}
    \label{fig:N1817nJuc23_VPD}
\end{figure}

\begin{figure}
	\includegraphics[width=\columnwidth]{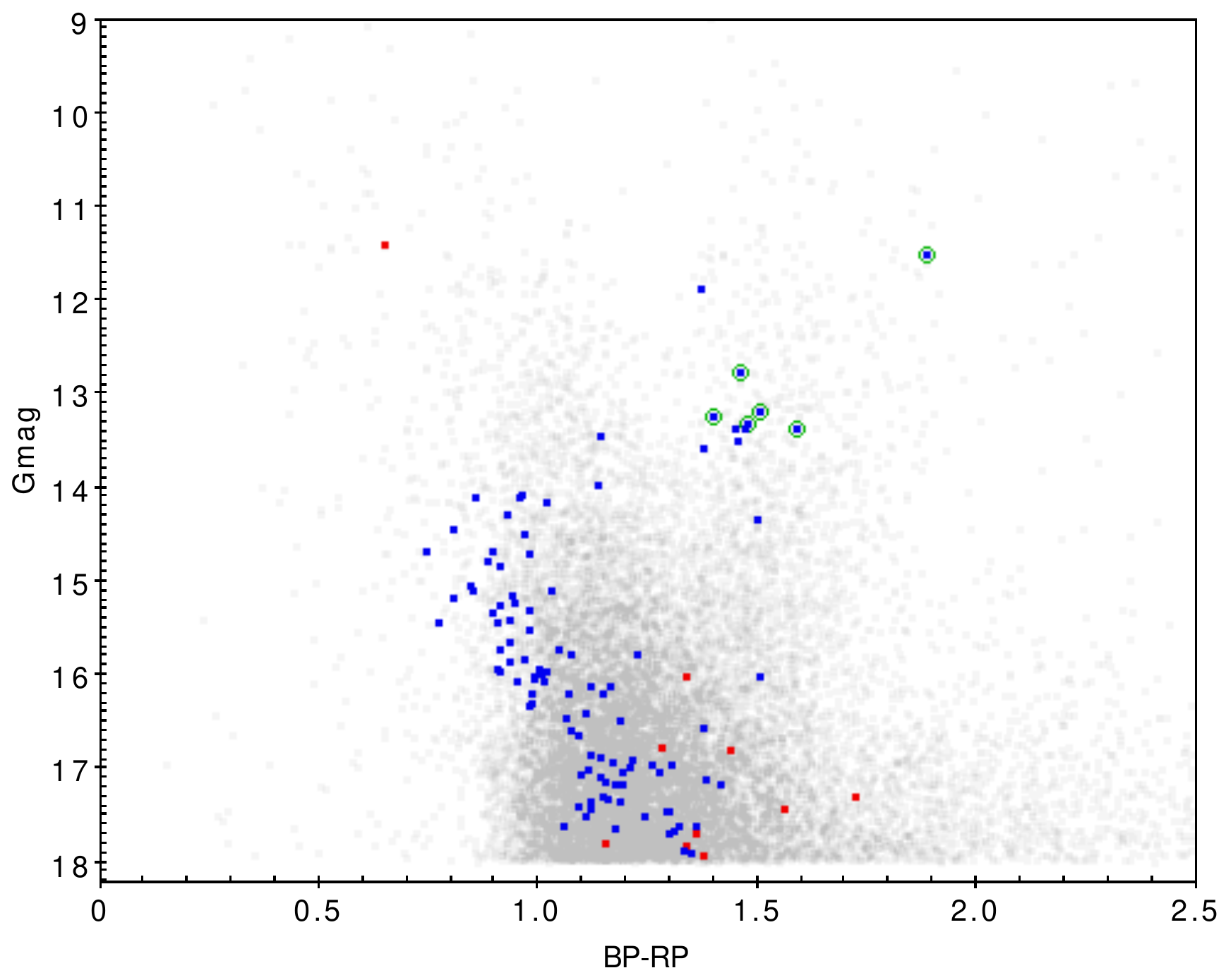}
    \caption{Colour Magnitude diagram of Juchert~23, in red and blue 115 members by \texttt{Clusterix} and in blue, the 105 members stars with an additional cut in parallax. In green open circles, 6 stars with coherent {\em Gaia} radial velocities.}
    \label{fig:HR_Juc23}
\end{figure}

After many tests in the area of NGC~1817, we were not able to find any hint for NGC~1807 asterism. However, we can confirm Juchert~23 (DSH~J0507.6+1734), a very small, little known cluster (see full area in Fig.~\ref{fig:N1817nJuc23_area} and the VPD of both in Fig.~\ref{fig:N1817nJuc23_VPD}. It is not in the lists of \citet{dias2002}, \citet{Kharchenko2016} or \citet{tristan2018b} and appears only in the list of suspected open cluster candidates from 2MASS and DSS in an additional Table~2e only in electronic form from \citet{Kronberger2006}. They only give coordinates and a visual diameter of 7$\arcmin$. We study an area of one degree radius around its centre with $G<$18 (15\,145 stars). \texttt{Clusterix} 2.0. gives us an estimated number of members of 115. With 105 members after a sigma cut in parallax, we get a clean Colour Magnitude diagram (Fig.~\ref{fig:HR_Juc23}).

The proper motion and parallax of the cluster are:
$\mu_{\alpha*}=$ 0.34$\pm$0.09\,mas yr$^{-1}$; $\mu_{\delta}=$ $-$2.56$\pm$0.09\,mas yr$^{-1}$; $\varpi=$ 0.35$\pm$0.09\,mas for 105 member stars. There are six stars with coherent {\em Gaia} radial velocity ($v_{\textnormal{rad}} = 0.5\pm$2.1\,km s$^{-1}$) that produce an average parallax of $\varpi=$ 0.34$\pm$0.04\,mas.


\subsection{Overlapping clusters: NGC~1750 and NGC~1758}
\label{NGC1750}

NGC~1750 and NGC~1758 are a pair of partly overlapping open clusters at around 700~pc with $E(B-V) =0.34$ \citep{Straicys1998b, Galadi1998, Galadi1998a, Galadi1998b, Tian1998}. 

\begin{figure}
	\includegraphics[width=\columnwidth]{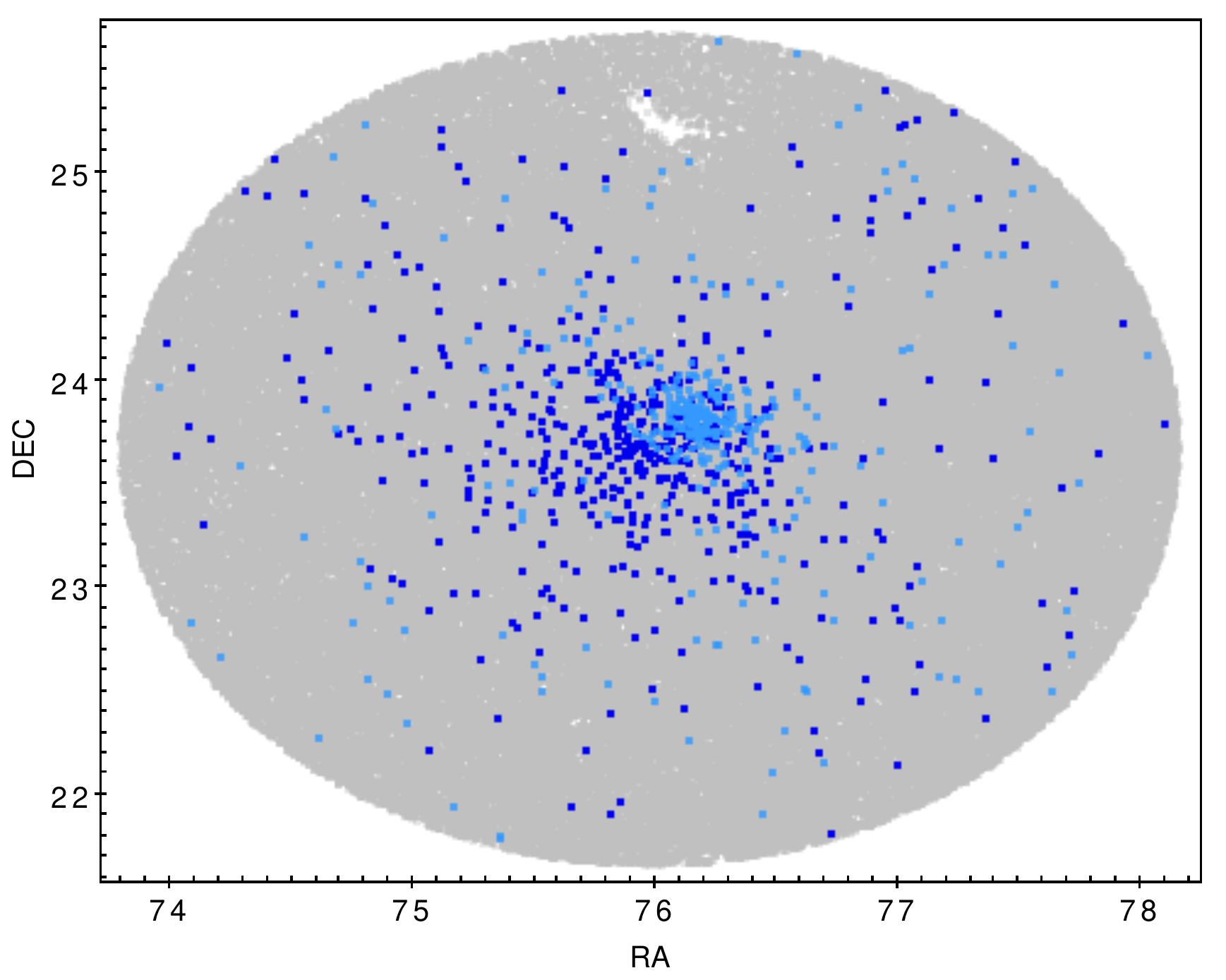}
    \caption{The two known overlapping clusters: NGC~1750 in blue with 504 member stars, and NGC~1758 in light blue with 363 member stars. In grey dots all stars in a 2 degree radius around the centre. }
    \label{fig:NGC175058_area}
\end{figure}

\begin{figure}
	\includegraphics[width=\columnwidth]{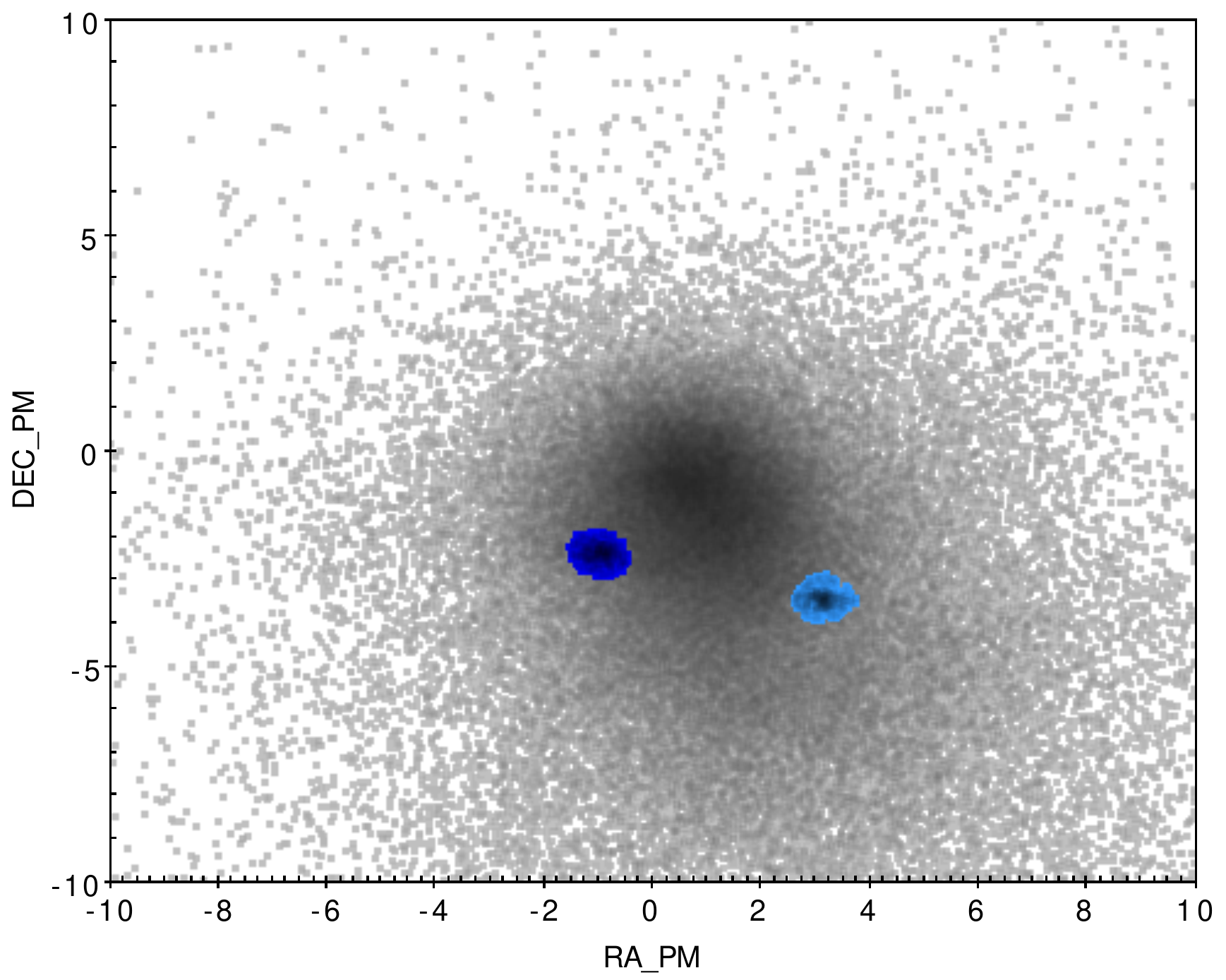}
    \caption{Vector point diagram of the two overlapping clusters: NGC~1750 in blue (504 stars) and NGC~1758 in light blue (363 stars). In gray all stars in a 2 degree radius around the centre. }
    \label{fig:NGC175058_VPD}
\end{figure}

\begin{figure}
	\includegraphics[width=\columnwidth]{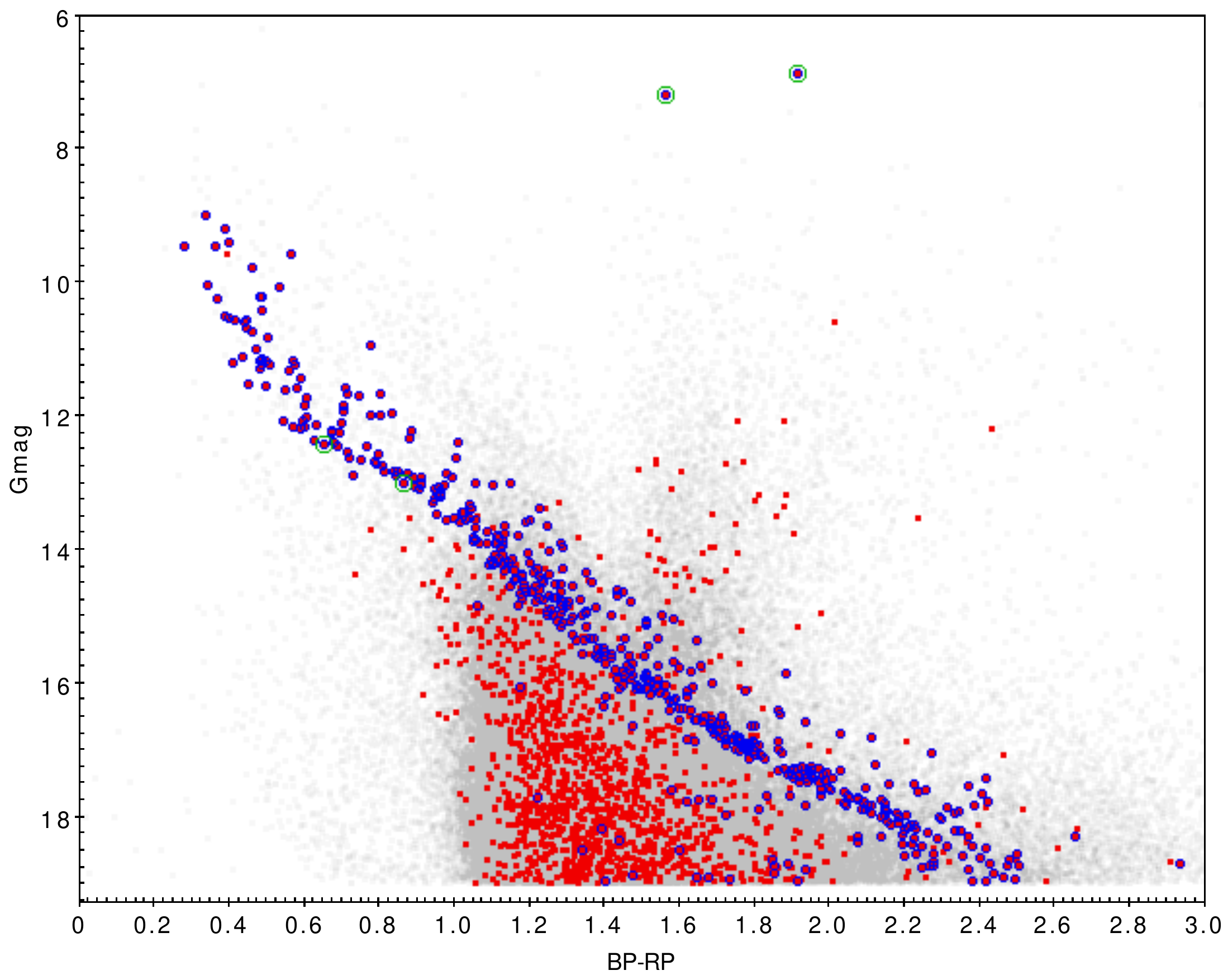}
    \caption{Colour Magnitude diagram of NGC~1750: in grey the 111\,799 stars in the field, in red and blue the 1827 candidates, and in blue the final selection of 504 stars. In green open circle the four stars with coherent {\em Gaia} radial velocities. See text for more details.}
    \label{fig:NGC1750_HR}
\end{figure}

\begin{figure}
	\includegraphics[width=\columnwidth]{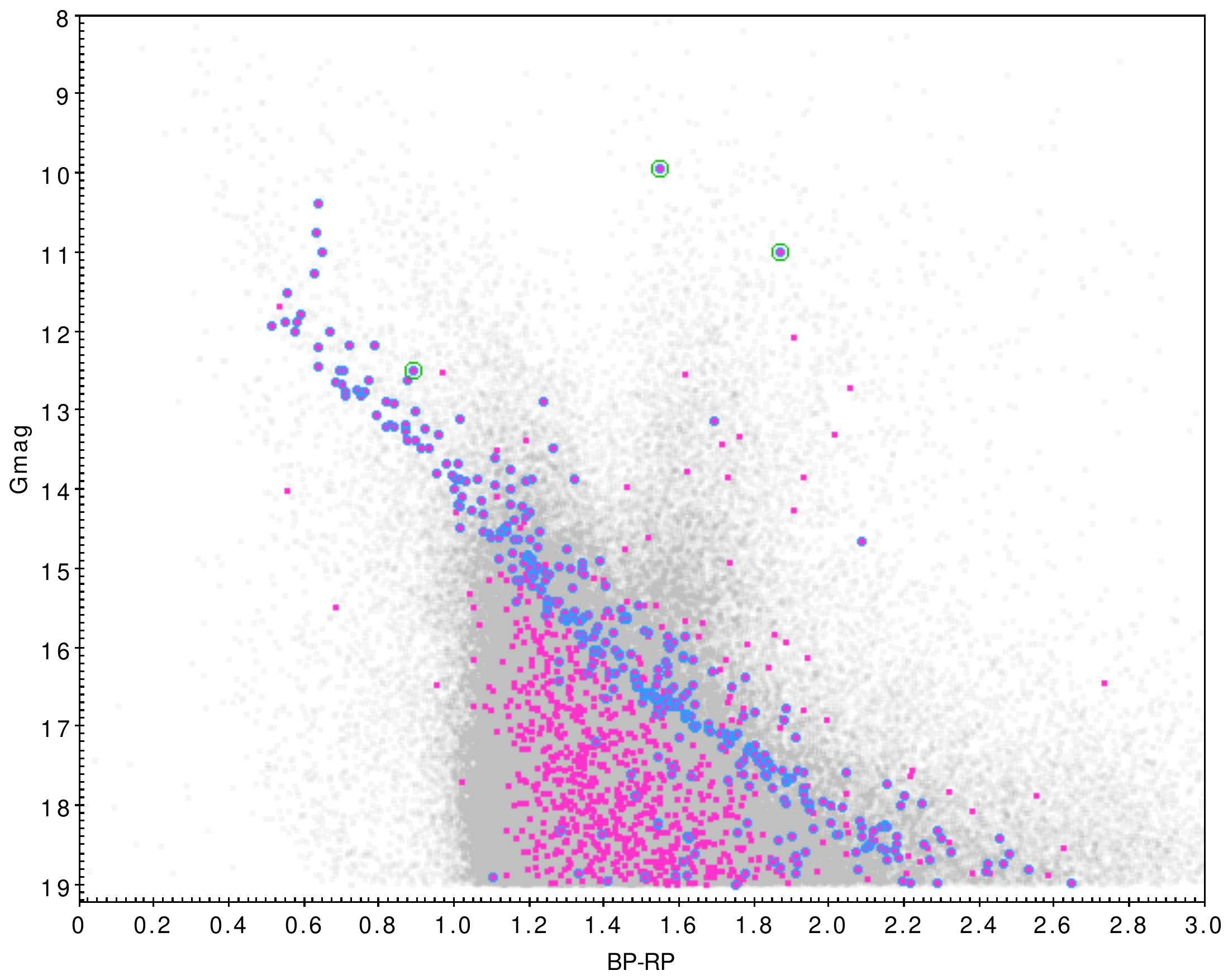}
    \caption{Colour Magnitude diagram of NGC~1758: in grey the 111\,799 stars in the field, in pink and light blue the 1162 candidates, and in light blue the final selection of 363 stars. In green open circle the three stars with coherent {\em Gaia} radial velocities. See text for more details.}
    \label{fig:NGC1758_HR}
\end{figure}

In an area of 2\,$\deg$ around NGC~1750 (111\,799 stars with $G<$19, see Fig.\ref{fig:NGC175058_area}) \texttt{Clusterix} finds two clusters with 2\,989 expected members in total. With these results we plot the proper motion diagram and we can clearly separate the two populations: NGC~1750 with 1\,827 candidate members and NGC~1758 with 1\,162 (see Fig.~\ref{fig:NGC175058_VPD}). By simply introducing a further cut in parallaxes, one can easily clean the colour-magnitude diagrams (see Figs.~\ref{fig:NGC1750_HR} and ~\ref{fig:NGC1758_HR}). The average proper motion of the clusters is:
$\mu_{\alpha*} = -0.97\pm0.23$\,mas yr$^{-1}$ ; $\mu_{\delta} = -2.40\pm0.21$\,mas yr$^{-1}$; $\varpi = 1.34\pm0.16$\,mas for 504 member stars in NGC~1750. There is only 4 stars with {\em Gaia} coherent radial velocities ($v_{\textnormal{rad}} = -11.5\pm0.7$\,km s$^{-1}$). Regarding NGC~1758 we obtain 
$\mu_{\alpha*}=$ 3.13$\pm0.23$\,mas yr$^{-1}$; $\mu_{\delta} = -3.45\pm0.20$\,mas yr$^{-1}$; $\varpi = 1.08\pm0.15$\,mas for 363 member stars. There is only 3 stars with coherent {\em Gaia} radial velocities ($v_{\textnormal{rad}} = 7.2\pm3.4$\,km s$^{-1}$). 

Comparing this result with the  member stars found in the study on the 1.5\,$\deg$ area with a magnitude limit of $G=18$ from \citet{tristan2018b}, we find 378 member stars in common for NGC~1750 (86$\%$ agreement) and 146 member stars in common for NGC~1758 (99$\%$ agreement). The comparison of the final values are in Table~\ref{tab:N175058}.

\begin{table*}
 \caption{Comparison of our results for NGC~1750 and NGC~1758 with other authors.}
 \label{tab:N175058}
 \begin{tabular}{llllll}
  \hline
   & $\mu_{\alpha*}$ & $\mu_{\delta}$ & $\varpi$ 
   & $N_{\textnormal{mem}}$ & $G_{\rm lim}$\\
      & (mas yr$^{-1}$) & (mas yr$^{-1}$) & (mas) & & (mag) \\
  \hline
  {\bf NGC~1750} & & & & &\\
  This study & $-$0.97$\pm$0.23 & $-$2.40$\pm$0.21 
  &  1.34$\pm$0.16 & 504 & 19 \\
  \citet{tristan2018b} & $-$0.96$\pm$0.246 & $-$2.366$\pm$0.201  & 1.361$\pm$0.09 & 439 & 18\\ 
 \hline
 {\bf NGC~1758} & & & & & \\
  This study & 3.13$\pm$0.23 & $-$3.45$\pm$0.20 
  &  1.08$\pm$0.15 & 363 & 19 \\
  \citet{tristan2018b} & 3.156$\pm$0.146 & $-$3.465$\pm$0.129  & 1.103$\pm$0.059 & 146 & 18\\ 
  \hline
 \end{tabular}
\end{table*}

\subsection{Ruprecht~26 and Clusterix~1}
\label{Rup26}

\begin{figure}
	\includegraphics[width=\columnwidth]{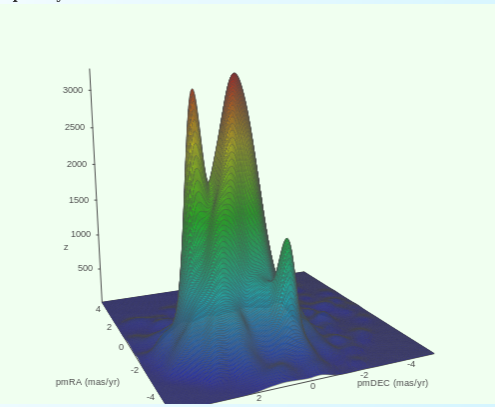}
    \caption{Frequency function of Ruprecht~26 area with the three kinematically differentiated populations: two clusters and the field in the middle.}
    \label{fig:canvas_Rup26.png}
\end{figure}

\begin{figure}
	\includegraphics[width=\columnwidth]{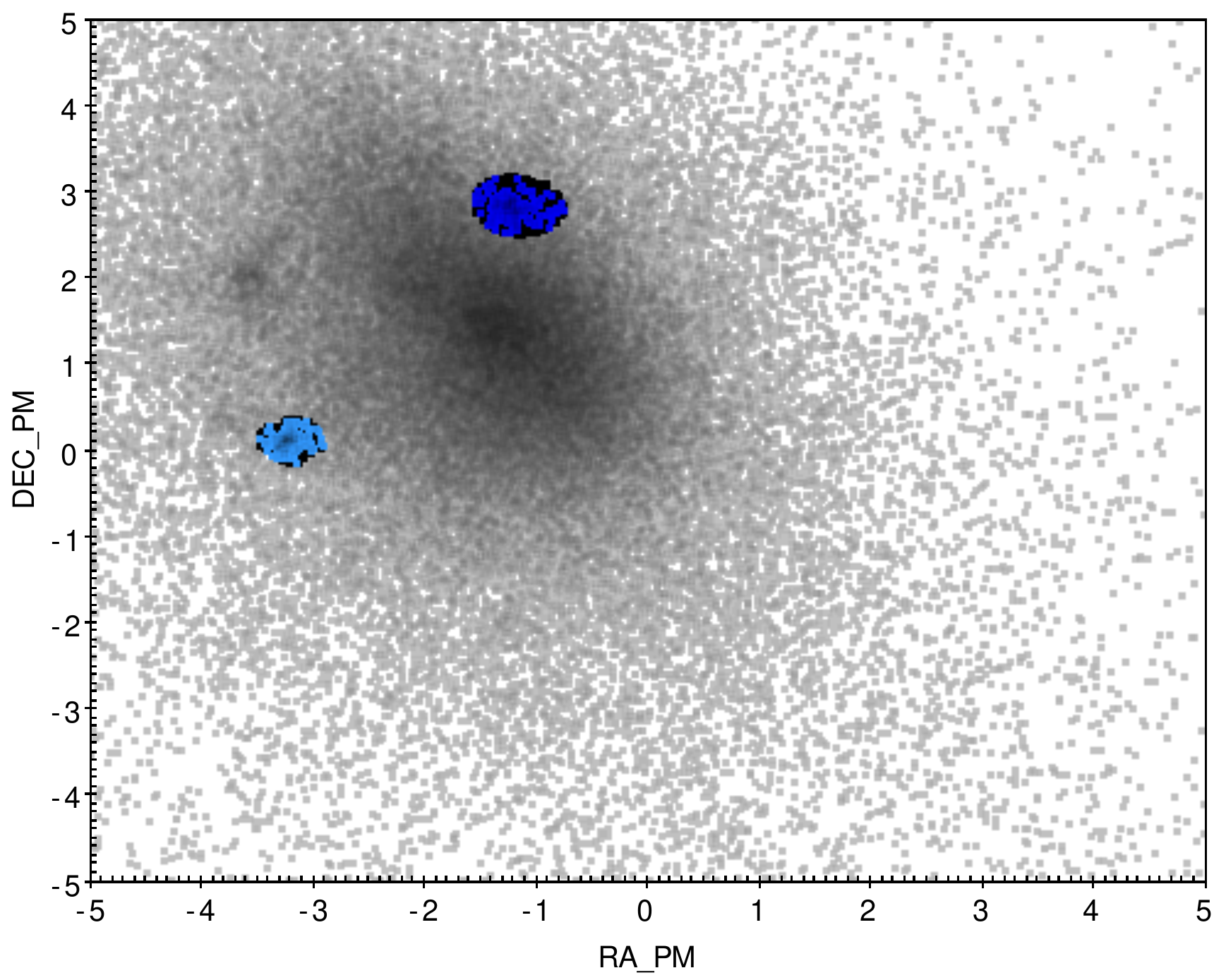}
    \caption{Vector point diagram of Ruprecht~26 (in light blue) and Clusterix~1 (in blue). In black all stars with probability greater than 0.5. In gray all stars in a 1 degree radius around the centre.}
    \label{fig:Rup26_VPD}
\end{figure}

In order to check areas with multiple clusters we have as well evaluated an area of one degree around Ruprecht~26, where many other known clusters exist (NGC~2428, NGC~2425, NGC~2414, Ruprecht~151 and Alessi~17). In the central area of Ruprecht~26 we have found two kinematically distinct populations, besides the field one, as can be seen in Fig.~\ref{fig:canvas_Rup26.png} and Fig.~\ref{fig:Rup26_VPD}. One of these populations corresponds to Ruprecht~26, and the other does not correspond to any of the previously mentioned known clusters.
 
In cases like this, where the field is full with other clusters, it is difficult to find a patch of sky mostly free of clusters. Nevertheless, after different trials we have managed to find a reasonable trade off. You can see the parameters used in the Appendix. In this case, due to the low constrast of the populations in the spatial plane, the expected number of cluster members is not reliable and we need to apply other criteria (photometry, parallax, spatial distribution, etc...) to clean the membership from proper motions. We accept as tentative members for both clusters all stars with membership probabilities higher than 0.5 (1\,569 candidate stars). 
 
\begin{figure}
	\includegraphics[width=\columnwidth]{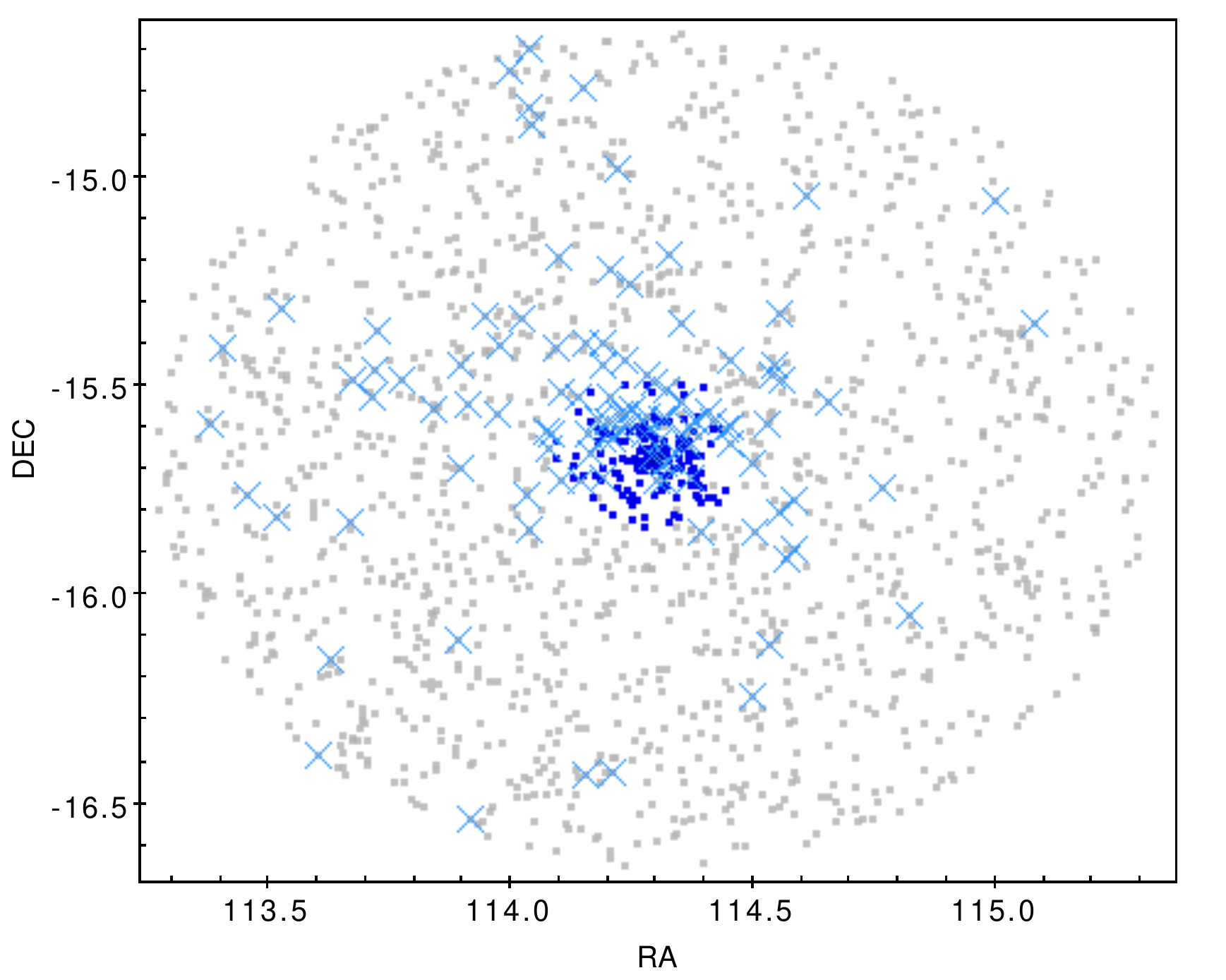}
    \caption{Area of Ruprecht~26 in light blue crosses, and Clusterix~1 in blue dots. In gray dots all stars with probability greater than 0.5. }
    \label{fig:Rup26nClusterix1_area}
\end{figure} 
 
The two populations found are superposed in the space; while one population with 1238 stars (see Fig.~\ref{fig:Rup26nClusterix1_area}) is very well concentrated, the other with 331 stars (Ruprecht~26) is more disperse and poorer. The differentiated population (that we called Clusterix~1 from now on) has only been discovered thanks to the exquisite quality of the {\em Gaia}~DR2 proper motions. To confirm their separate existence we study their Colour Magnitude diagrams and parallax distribution. 

\begin{figure}
	\includegraphics[width=\columnwidth]{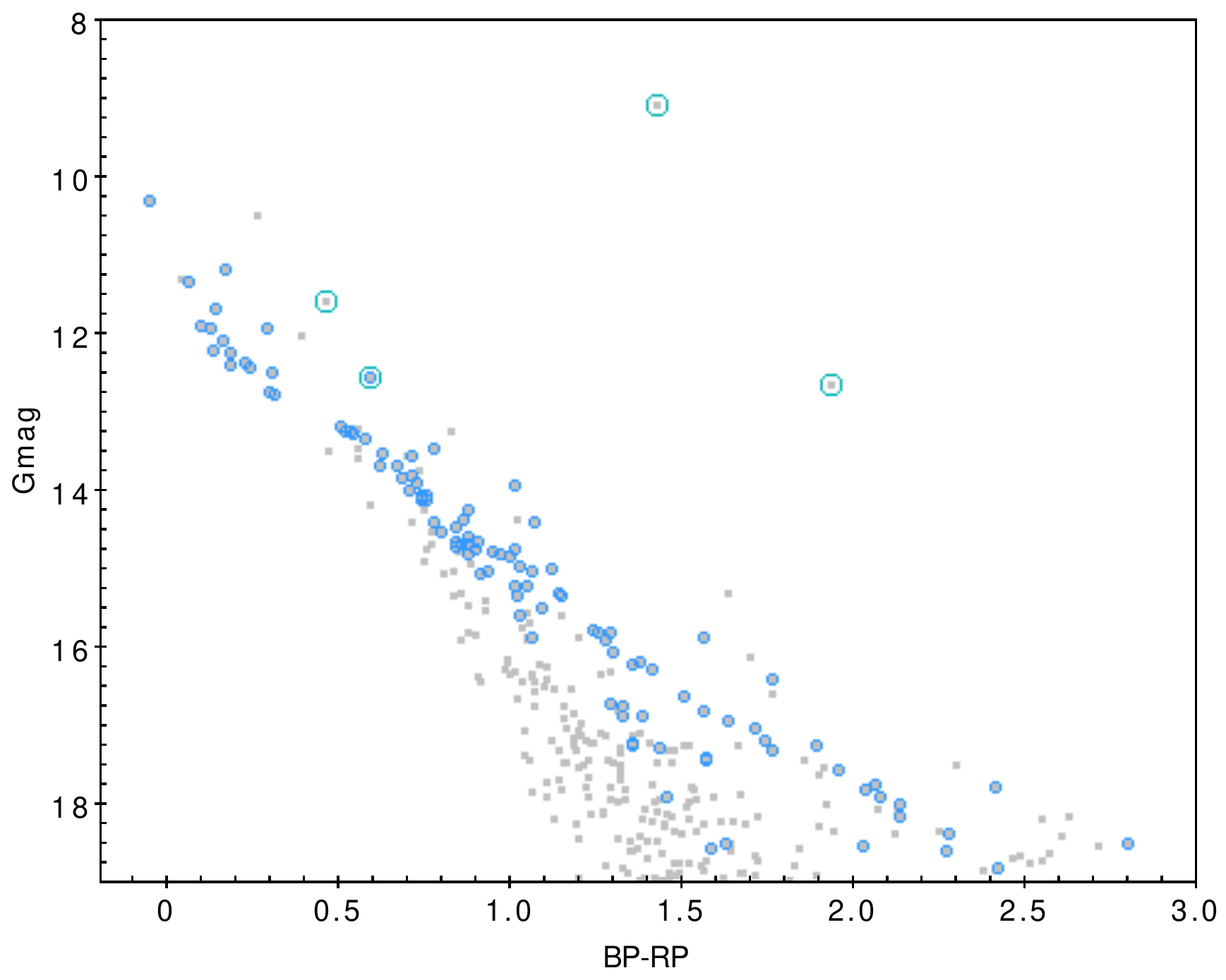}
    \caption{Colour Magnitude diagram of Ruprecht~26. The 109 members in light blue with 4 stars with coherent {\em Gaia} radial velocities in open green circles. In gray the 331 candidate stars. See more details in the text. }
    \label{fig:Rup26_phot}
\end{figure}

Ruprecht~26 candidate members show a distinct parallax so, applying a cut selection (0.8$<\varpi<$1.1\,mas) we can see a clear main sequence (see Fig.~\ref{fig:Rup26_phot}). From this final selection of 109 members we find an average proper motion of 
$\mu_{\alpha*} = -3.20\pm0.12$\,mas yr$^{-1}$ ; $\mu_{\delta} = 0.11\pm0.10$\,mas yr$^{-1}$ and $\varpi = 0.92\pm0.06$\,mas. There is only four stars with coherent {\em Gaia} radial velocities $v_{\textnormal{rad}} = 44.8\pm4.8$\,km s$^{-1}$.

\begin{figure}
	\includegraphics[width=\columnwidth]{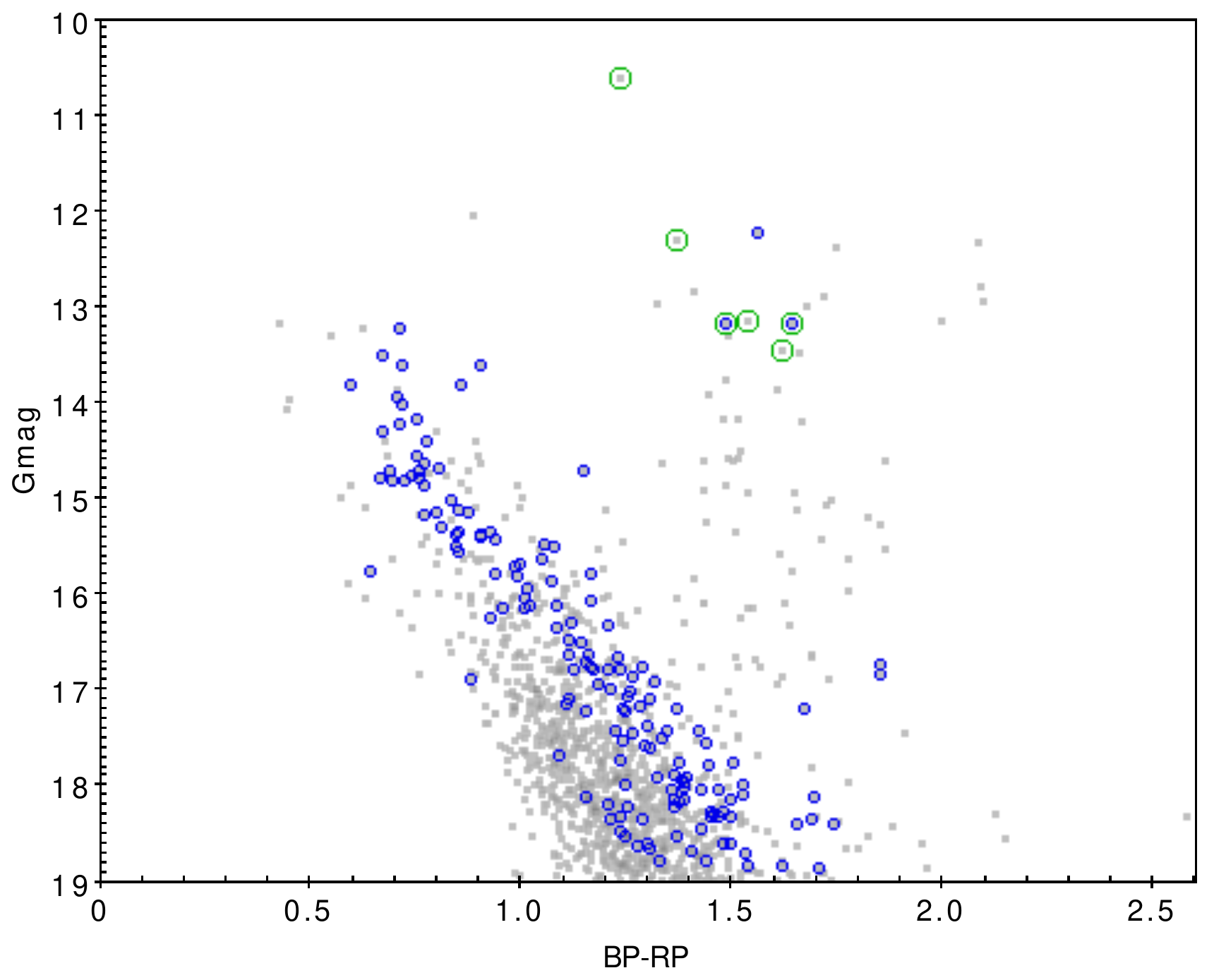}
    \caption{Colour Magnitude diagram of Clusterix~1. The 157 members in blue with 6 stars with coherent {\em Gaia} radial velocities in open green circles. In grey the 1\,238 candidate stars. See more details in the text. }
    \label{fig:Clus_phot}
\end{figure}

In the case of Clusterix~1, its parallax is small and it is difficult to separate from the field, so we do not use parallaxes and just select a core area ($r$ $<$ 0.2 deg) to see a cleaner distribution in the Colour Magnitude diagram (see Fig.\ref{fig:Clus_phot}). With that selection we found an average proper motion of $\mu_{\alpha*}=$ $-1.20\pm0.16$\,mas yr$^{-1}$ ; $\mu_{\delta}=$ 2.80$\pm$0.13\,mas yr$^{-1}$ from 157 members and a parallax of $0.41\pm0.15$\,mas. With these parameters, we find 6 stars with coherent {\em Gaia} radial velocities in the whole area, giving a $v_{\textnormal{rad}}=$ 37.3$\pm$5.2\,km s$^{-1}$). (three of them inside the core area).

Comparing Ruprecht~26 result with the  member stars found in the study on the 1\,$\deg$ area with a magnitude limit of $G=18$ by \citet{tristan2018b}, we find 69 member stars in common (85$\%$ agreement). The comparison of the final values are in Table~\ref{tab:Rup26}.

\begin{table*}
 \caption{Comparison of our results for Rup~26 with other authors.}
 \label{tab:Rup26}
 \begin{tabular}{llllll}
  \hline
   & $\mu_{\alpha*}$ & $\mu_{\delta}$ & $\varpi$ 
   & $N_{\textnormal{mem}}$ & $G_{\rm lim}$ \\
      & (mas yr$^{-1}$) & (mas yr$^{-1}$) & (mas) & & (mag) \\
  \hline
  This study & $-$3.20$\pm$0.12 & 0.11$\pm$0.10 
  &  0.92$\pm$0.06 & 109 & 19 \\
  \citet{tristan2018b} & $-$3.2$\pm$0.094 & 0.071$\pm$0.092  & 0.923$\pm$0.049 & 81 & 18 \\
  \hline
 \end{tabular}
\end{table*}



\begin{table*}
 \caption{Summary of mean parameters for the OCs characterised in this study (the full table is available in electronic version only.}
 \label{tab:generalresults}
 \begin{tabular}{lllll}
  \hline

 \end{tabular}
\end{table*}  

\section{Conclusions}
\label{conclusions}

Many automatic tools have been developed in recent years to separate cluster members from the field. We present here an open access web tool, VO compliant, to facilitate membership studies from proper motions data to any user that requires a tailor-made study on any specific data set. Different tools are able to apply to different cases and all of them have been able to discover new clusters. The census of Open Cluster seems far from complete.

We present the first results of \texttt{Clusterix} 2.0 for different cases to show the capabilities of the tool, its flexibility and adaptability to different environments: an area of five degrees around NGC~2682 (M~67), an old, well known cluster; a young close-by cluster NGC~2516 with a striking elongate structure extended up to four degrees that deserve further studies; NGC~1750 \& NGC~1758, a pair of partly overlapping clusters found without applying any a priori knowledge; in the area of NGC~1817 we confirm the non existence of NGC~1807 and the existence of a little known cluster, Juchert~23; and in an area with many neighbouring clusters we are able to disentangle the existence of two clusters where only one was previously known: Ruprecht~26 and the new, Clusterix~1.

\section*{Acknowledgements}

This work has made use of data from the European Space Agency (ESA) mission
{\it Gaia} (\url{https://www.cosmos.esa.int/gaia}), processed by the {\it Gaia}
Data Processing and Analysis Consortium (DPAC,
\url{https://www.cosmos.esa.int/web/gaia/dpac/consortium}). Funding for the DPAC
has been provided by national institutions, in particular the institutions
participating in the {\it Gaia} Multilateral Agreement.
This work was partly supported  by the Spanish State Research Agency (AEI) projects ESP2017-87676-C5-1-R, ESP2016-80079-C2-1-R and RTI2018-095076-B-C21 (MINECO/FEDER, UE), and MDM-2014-0369 of ICCUB and MDM-2017-0737  Centro de Astrobiolog\'{\i}a (CSIC-INTA) (Unidades de Excelencia 'Mar\'{\i}a de Maeztu'); as well as the European Community's Seventh Framework Programme (FP7/2007-2013) under grant agreement GENIUS FP7 - 606740 and Horizon H2020 Framework Programme under grant agreement ASTERICS - 653477. This publication makes use of VOSA, developed under the Spanish Virtual Observatory project supported by the Spanish MINECO through grant AyA2017-84089.
This work has made extensive use of Topcat \citep{Taylor2005},
and of NASA's Astrophysics Data System.




\bibliographystyle{mnras}
\bibliography{Clusterix} 




\appendix

\section{\texttt{Clusterix} procedure for our science cases}

We present here as case examples some of the choices made in \texttt{Clusterix} to produce the results explained before. 

\begin{figure}
	\includegraphics[width=\columnwidth]{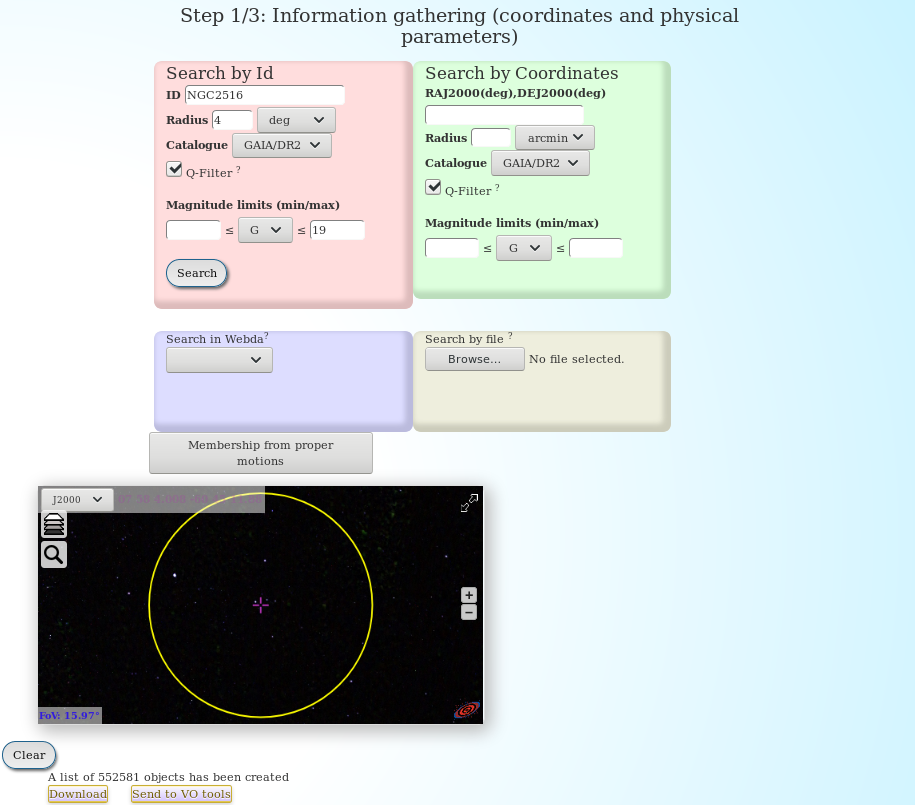}
    \caption{Step~1 of the \texttt{Clusterix} procedure for NGC~2516.}
    \label{fig:N2516_1}
\end{figure}

\begin{figure}
	\includegraphics[width=\columnwidth]{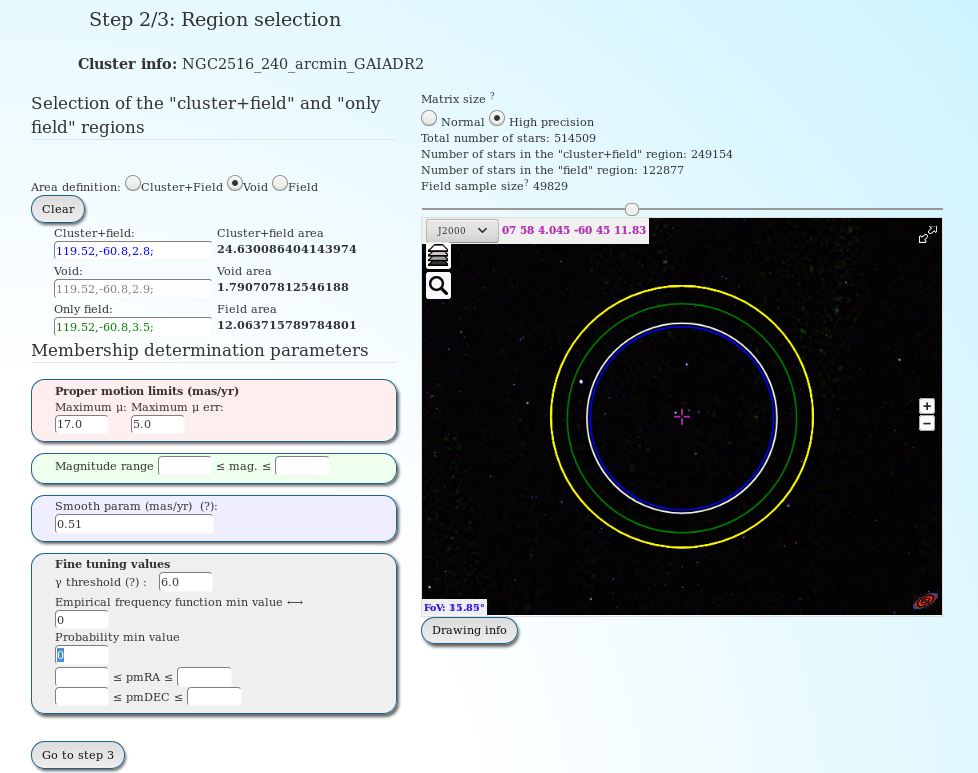}
    \caption{Step~2, in the case of NGC~2516 we have as well limited the number of stars in the field to around 50 thousand. High precision is used here for final results.}
    \label{fig:N2516_2a2}
\end{figure}

\begin{figure}
	\includegraphics[width=\columnwidth]{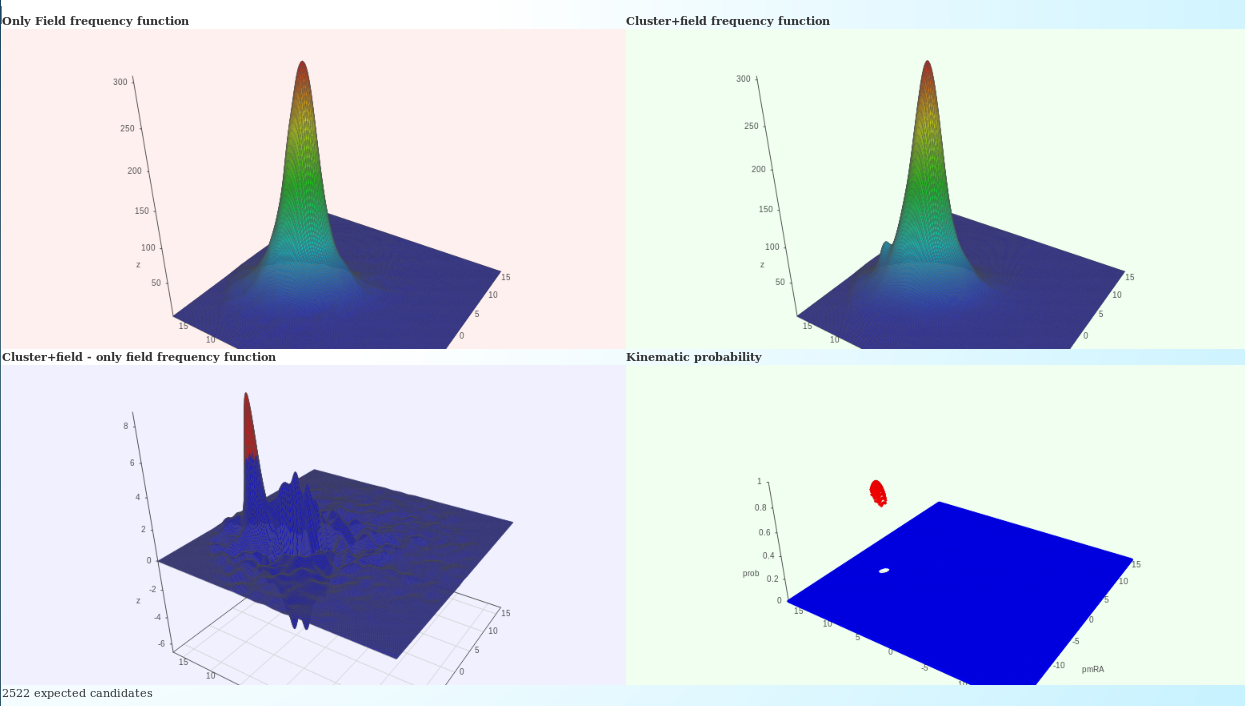}
    \caption{Interactive results of Step~2 of the \texttt{Clusterix} procedure for NGC~2516.}
    \label{fig:N2516_2b}
\end{figure}

\begin{figure}
	\includegraphics[width=\columnwidth]{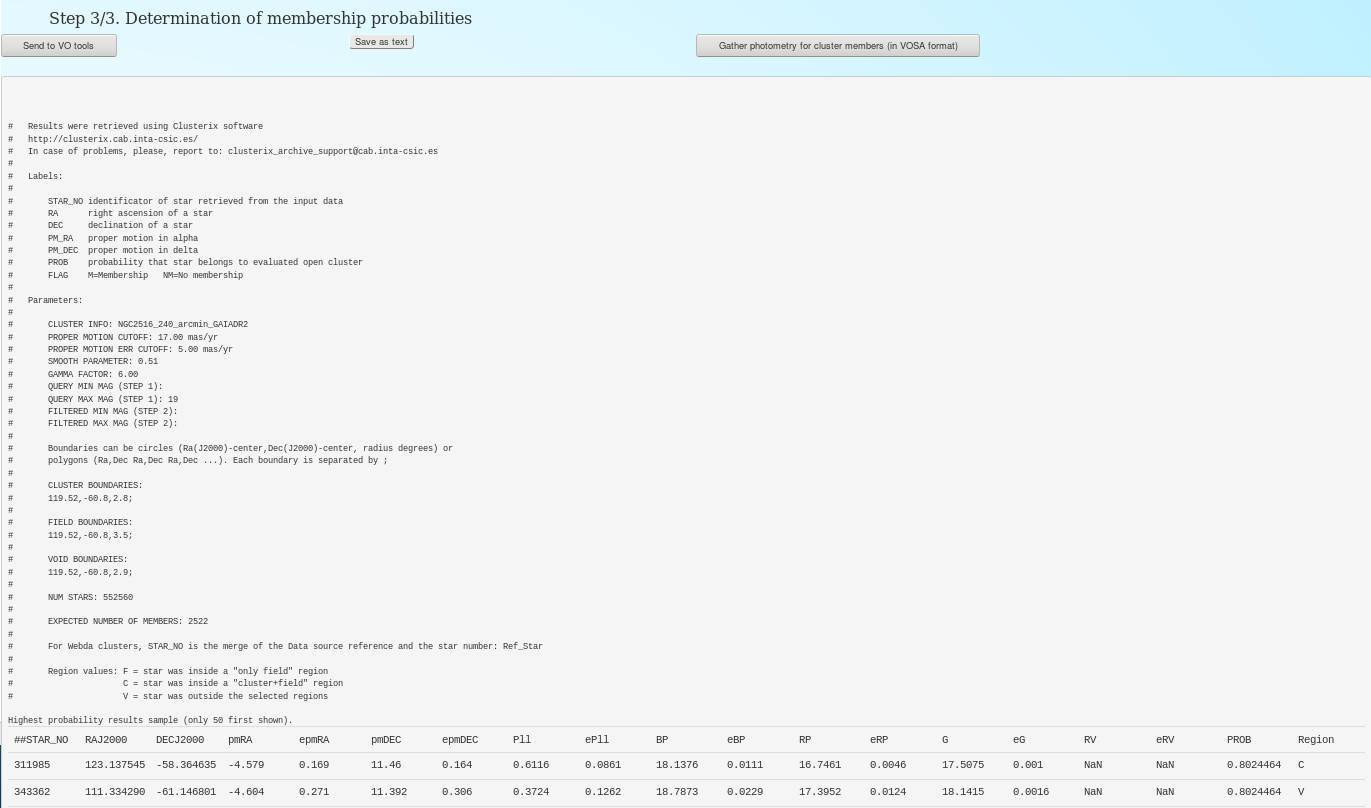}
    \caption{Step~3 of the \texttt{Clusterix} procedure for NGC~2516. In the file we can see the value of expected members is 2522, based on that value we look for the probability value corresponding to the 2522nd star and we found a value of 0.76047766, so we accept as candidate members all the stars with equal or greater probability, what gives us a value of 2537 candidate members.}
    \label{fig:N2516_3}
\end{figure}

For NGC~2516 see Figs.~\ref{fig:N2516_1},
\ref{fig:N2516_2a2},~\ref{fig:N2516_2b}, \ref{fig:N2516_3}.


\begin{figure}
	\includegraphics[width=\columnwidth]{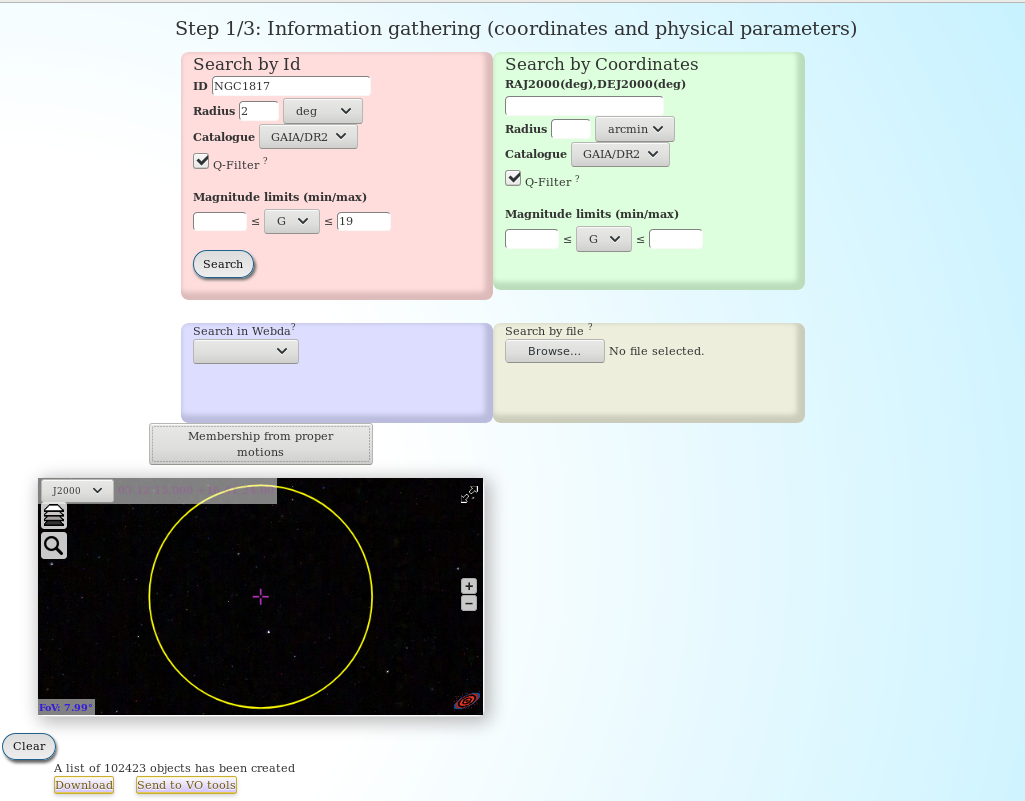}
    \caption{Step~1 of the \texttt{Clusterix} procedure for NGC~1817.}
    \label{fig:N1817_1}
\end{figure}

\begin{figure}
	\includegraphics[width=\columnwidth]{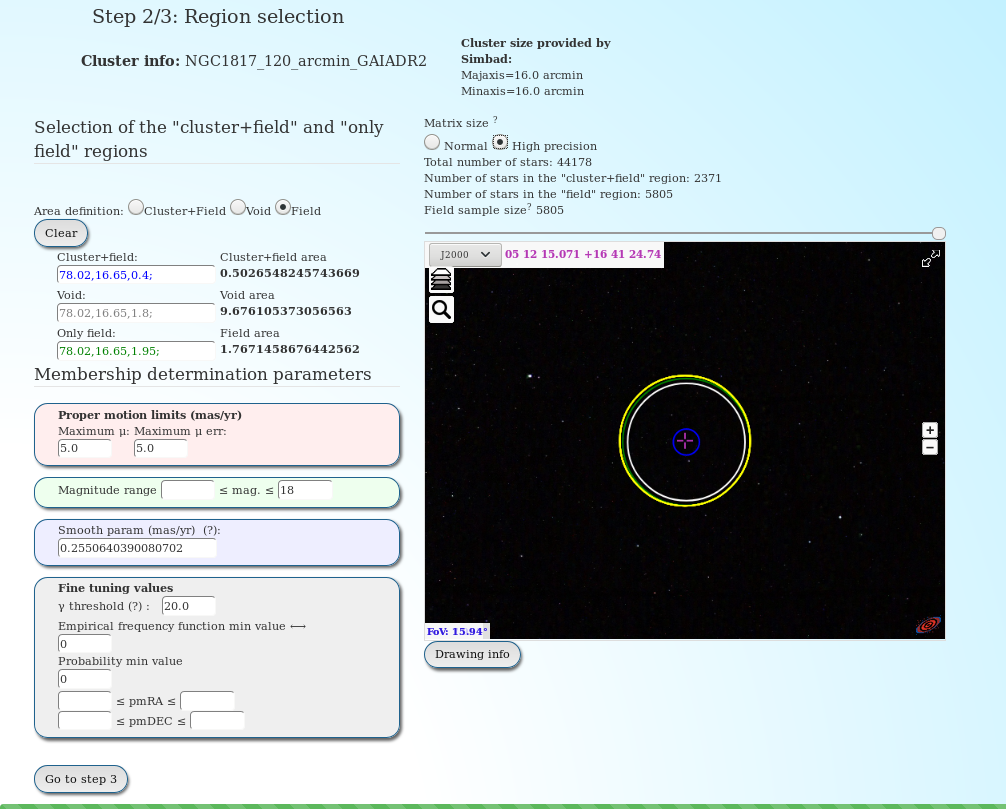}
    \caption{Step~2 where all the fine tuning parameters of the \texttt{Clusterix} procedure are set for NGC~1817.}
    \label{fig:N1817_2a}
\end{figure}

\begin{figure}
	\includegraphics[width=\columnwidth]{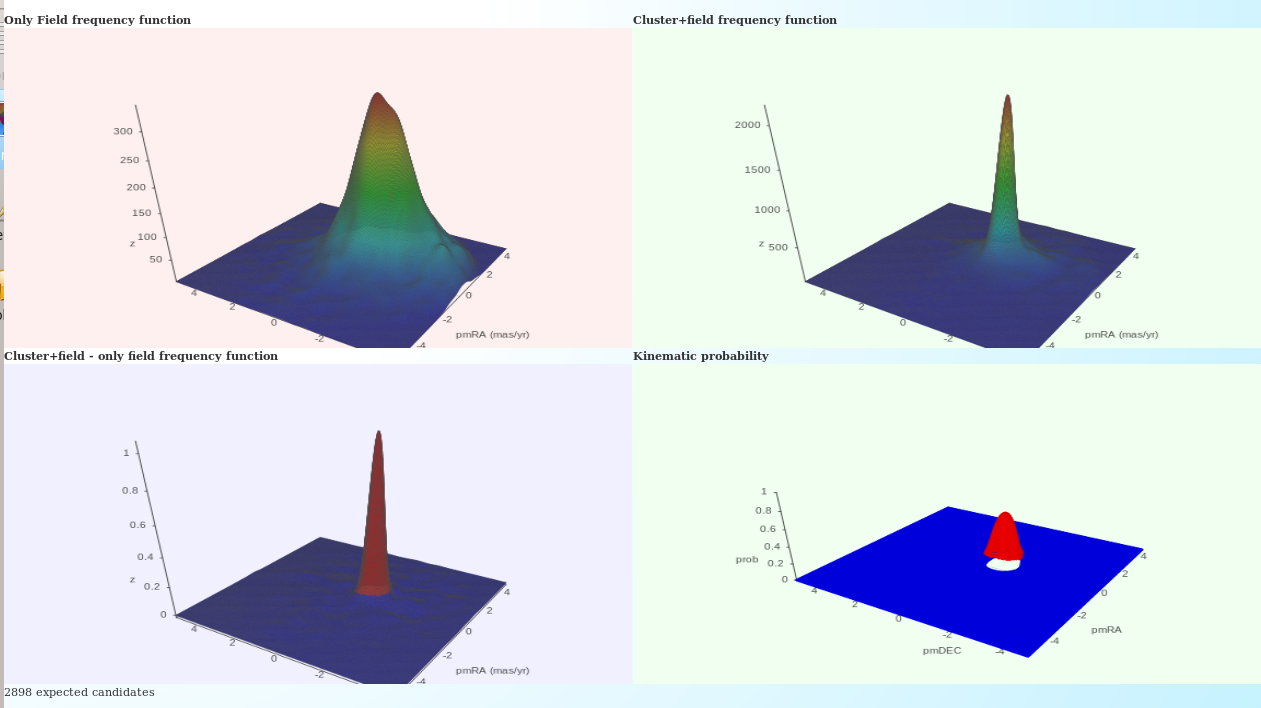}
    \caption{Interactive results of Step~2 of the \texttt{Clusterix} procedure for NGC~1817.}
    \label{fig:N1817_2b}
\end{figure}

\begin{figure}
	\includegraphics[width=\columnwidth]{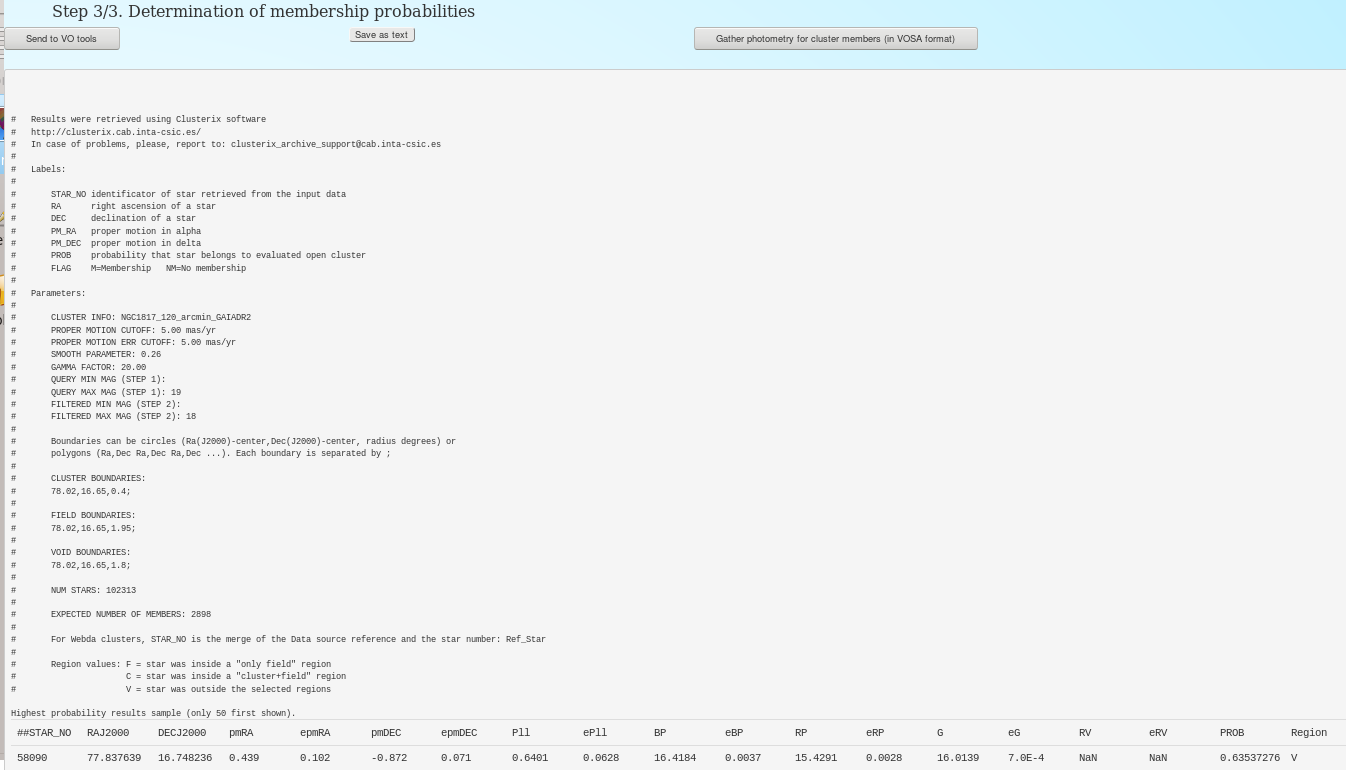}
    \caption{Step~3 of the \texttt{Clusterix} procedure for NGC~1817.}
    \label{fig:N1817_3}
\end{figure}

For NGC~1817 see Figs.~\ref{fig:N1817_1},~\ref{fig:N1817_2a},~\ref{fig:N1817_2b}, \ref{fig:N1817_3}.


\begin{figure}
	\includegraphics[width=\columnwidth]{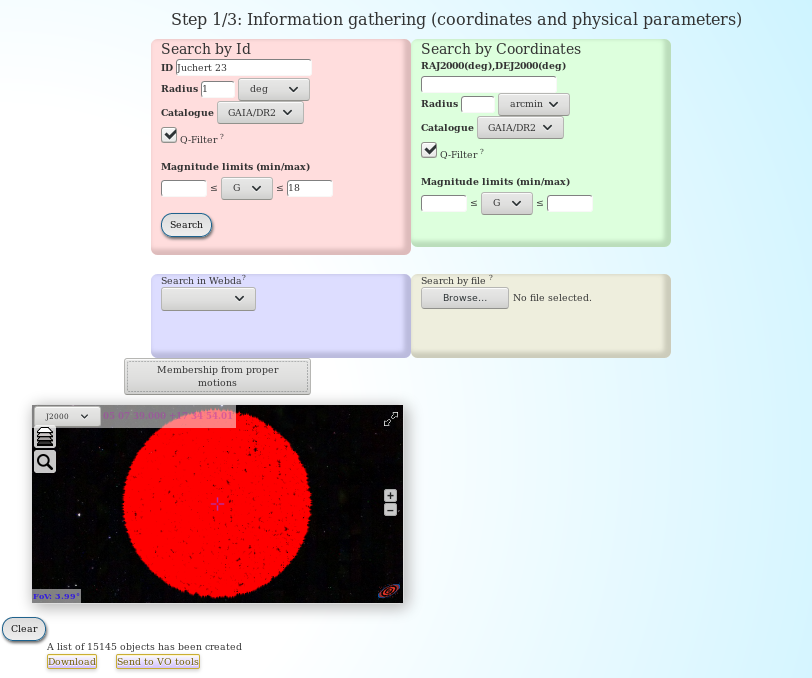}
    \caption{Step~1 of the \texttt{Clusterix} procedure for Juchert~23. We chose the database, the radius, the magnitude limit and can download the results of this search.}
    \label{fig:Ju23_1}
\end{figure}

\begin{figure}
	\includegraphics[width=\columnwidth]{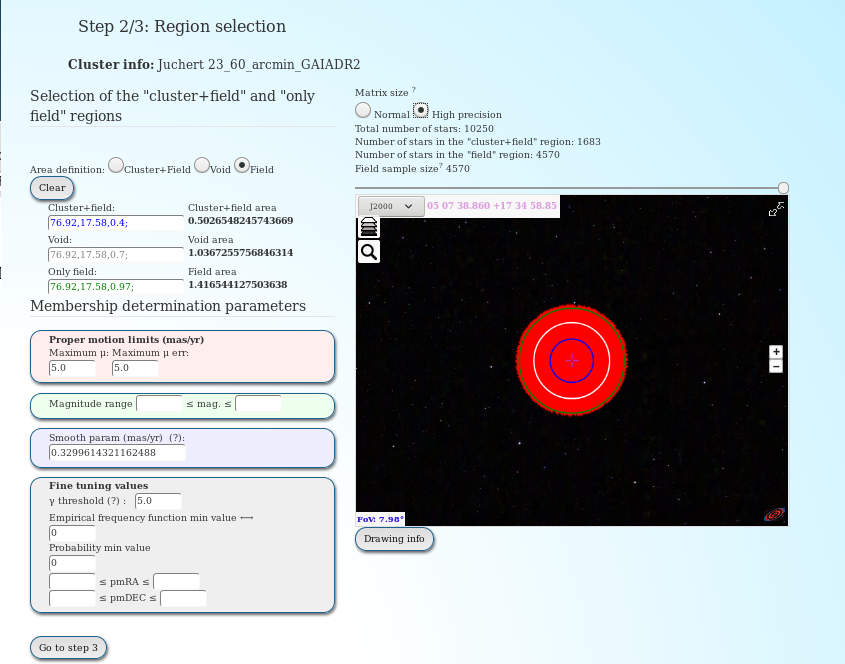}
    \caption{Step~2 where all the fine tuning parameters of the \texttt{Clusterix} procedure are set for Juchert~23.}
    \label{fig:Ju23_2a}
\end{figure}

\begin{figure}
	\includegraphics[width=\columnwidth]{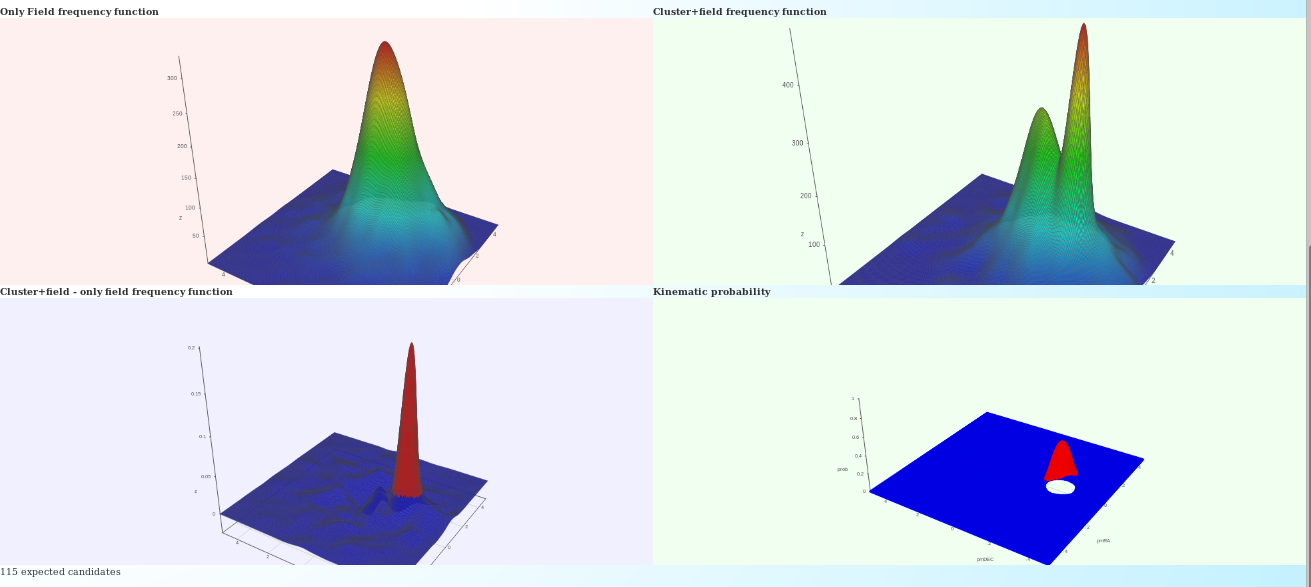}
    \caption{Interactive results of Step~2 of the \texttt{Clusterix} procedure for Juchert~23.}
    \label{fig:Ju23_2b}
\end{figure}

\begin{figure}
	\includegraphics[width=\columnwidth]{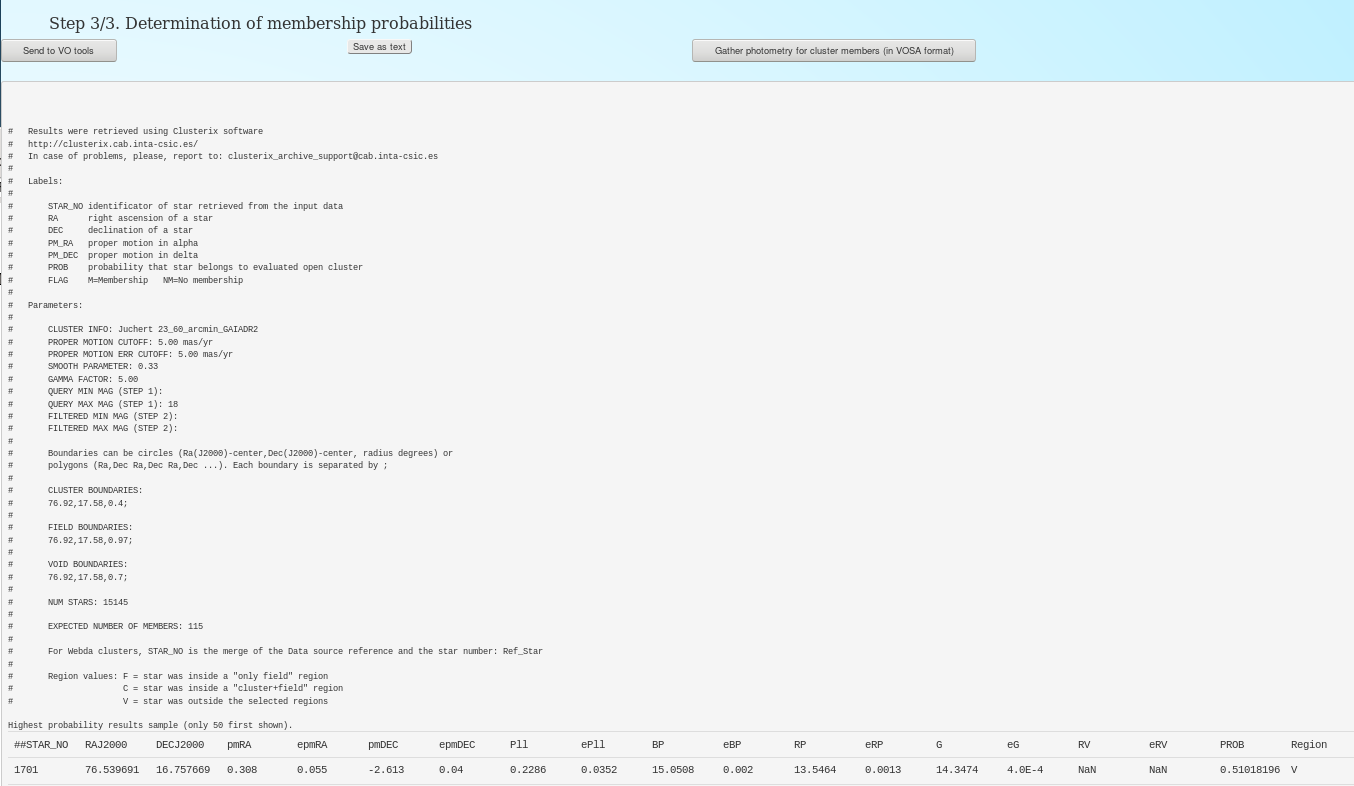}
    \caption{Step~3 of the \texttt{Clusterix} procedure for Juchert~23. }
    \label{fig:Ju23_3}
\end{figure}

For Juchert~23 see Figs.~\ref{fig:Ju23_1},~\ref{fig:Ju23_2a},~\ref{fig:Ju23_2b}, \ref{fig:Ju23_3}.


\begin{figure}
	\includegraphics[width=\columnwidth]{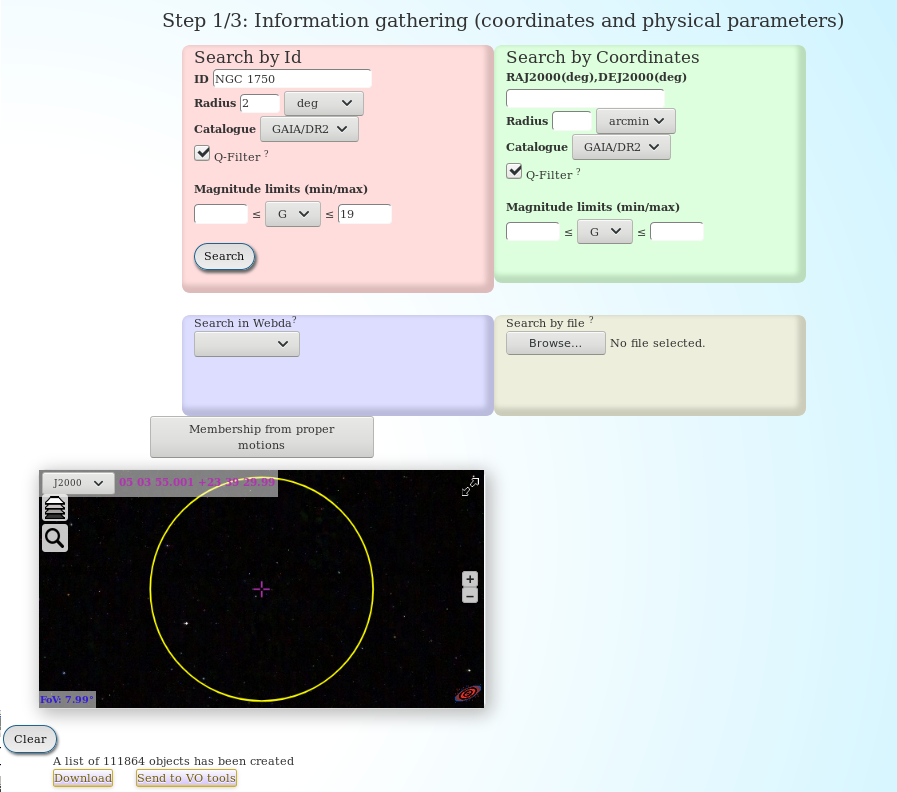}
    \caption{Step~1 of the \texttt{Clusterix} procedure for NGC~1750 and NGC~1758.}
    \label{fig:N175058_1}
\end{figure}

\begin{figure}
	\includegraphics[width=\columnwidth]{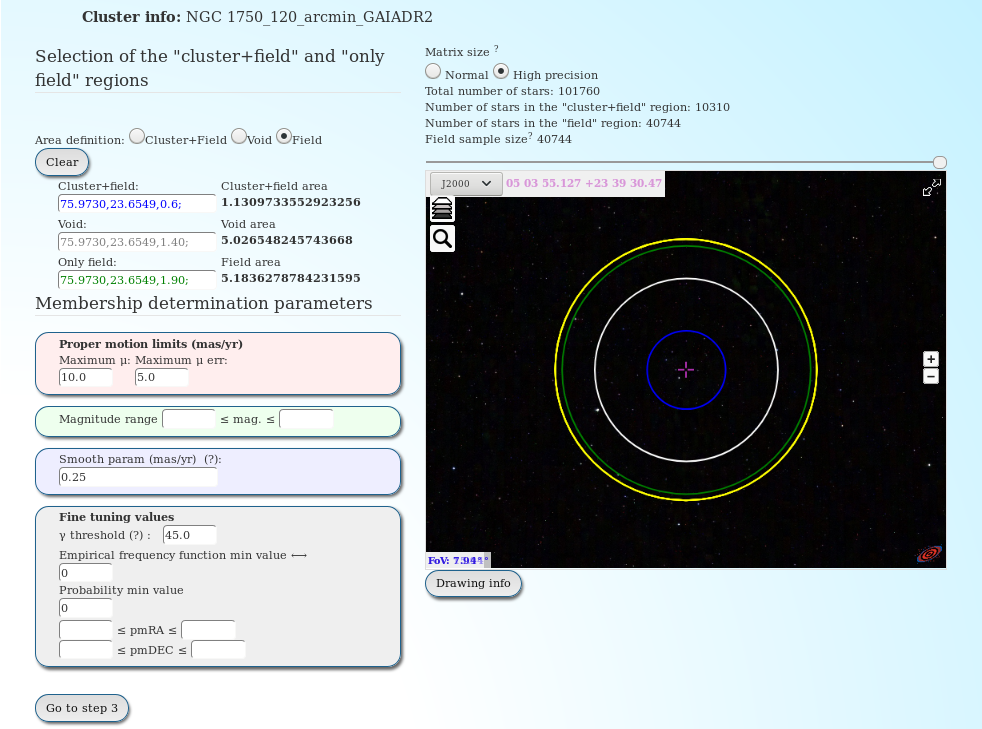}
    \caption{Step~2 where all the fine tuning parameters of the \texttt{Clusterix} procedure are set for NGC~1750 and NGC~1758.}
    \label{fig:N175058_2a}
\end{figure}

\begin{figure}
	\includegraphics[width=\columnwidth]{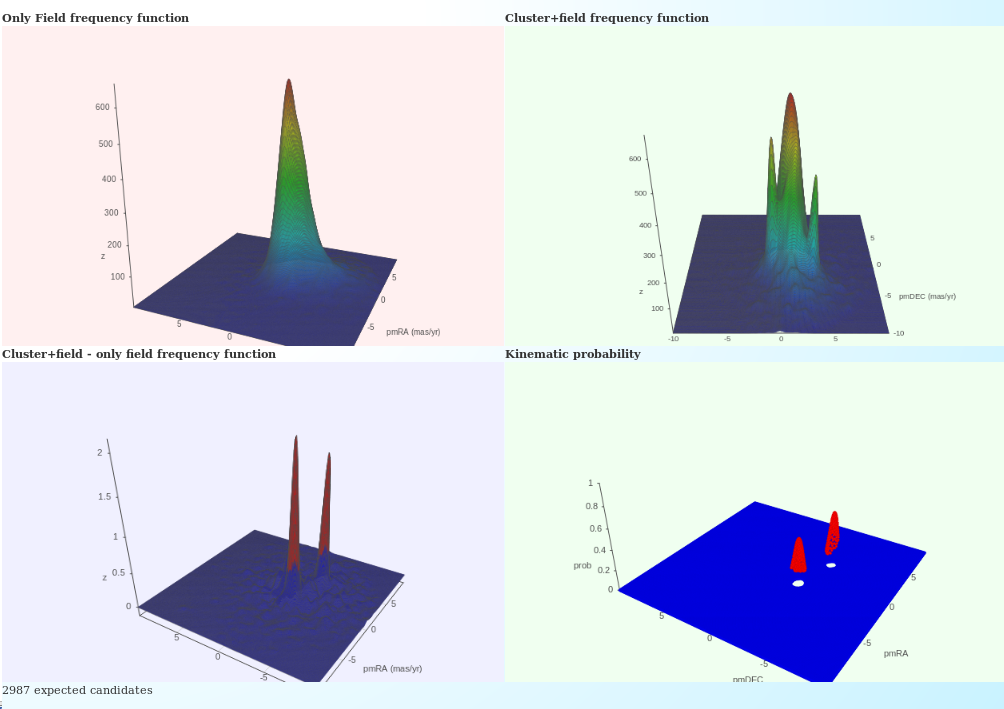}
    \caption{Interactive results of Step~2 of the \texttt{Clusterix} procedure for NGC~1750 and NGC~1758.}
    \label{fig:N175058_2b}
\end{figure}

\begin{figure}
	\includegraphics[width=\columnwidth]{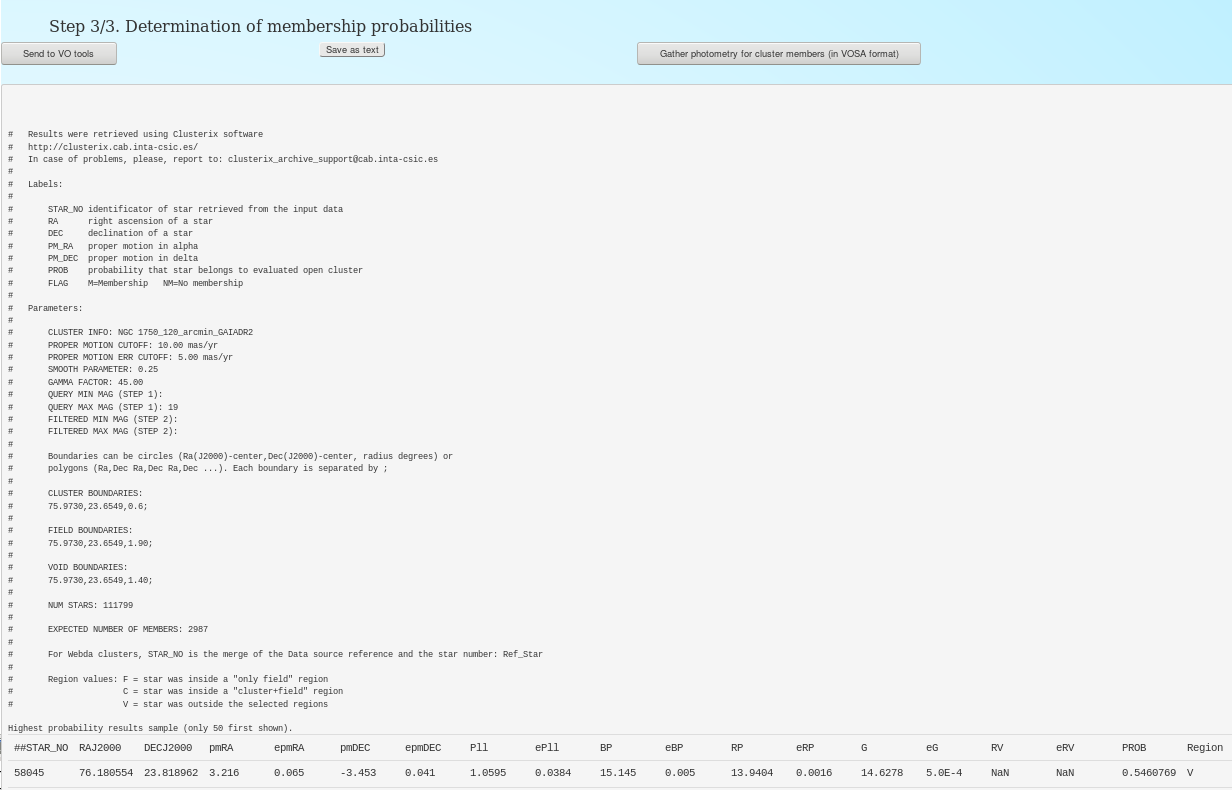}
    \caption{Step~3 of the \texttt{Clusterix} procedure for NGC~1750 and NGC~1758.}
    \label{fig:N175058_3}
\end{figure}

For NGC~1750 and NGC~1758 see Figs.~\ref{fig:N175058_1},~\ref{fig:N175058_2a},~\ref{fig:N175058_2b}, \ref{fig:N175058_3}.


\begin{figure}
	\includegraphics[width=\columnwidth]{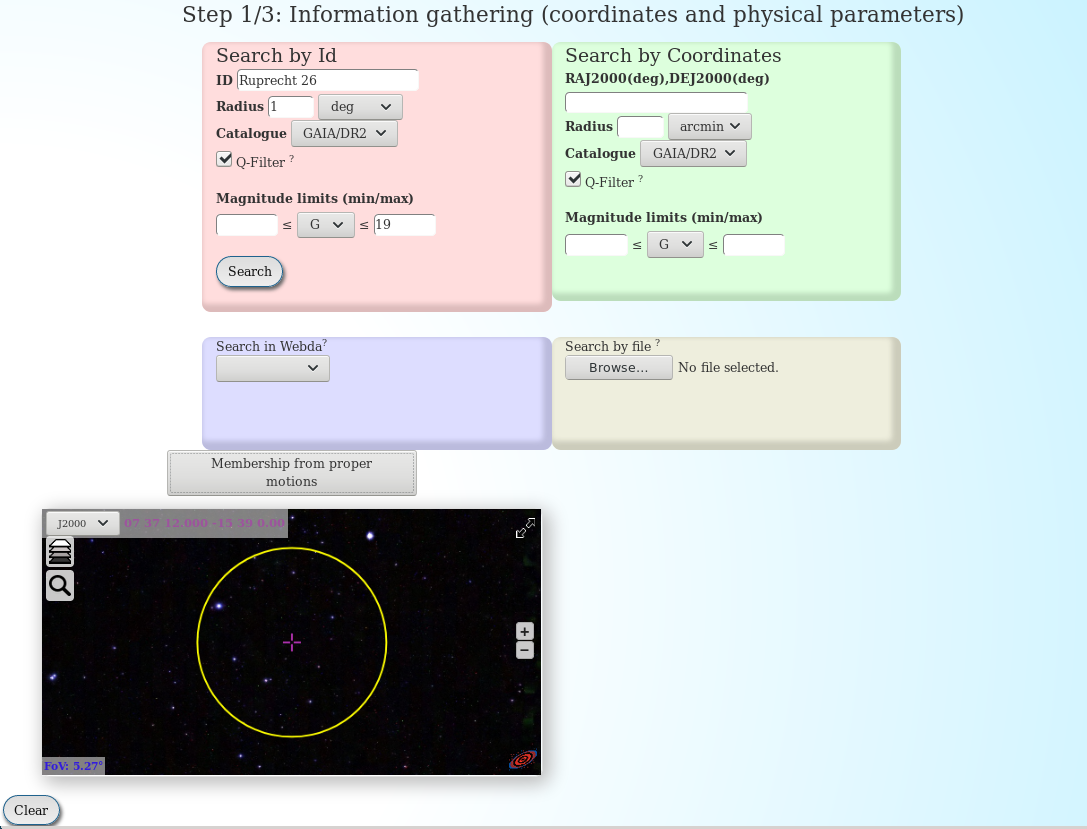}
    \caption{Step~1 of the \texttt{Clusterix} procedure for Ruprecht~26 and Clusterix~1.}
    \label{fig:Rup26_1}
\end{figure}

\begin{figure}
	\includegraphics[width=\columnwidth]{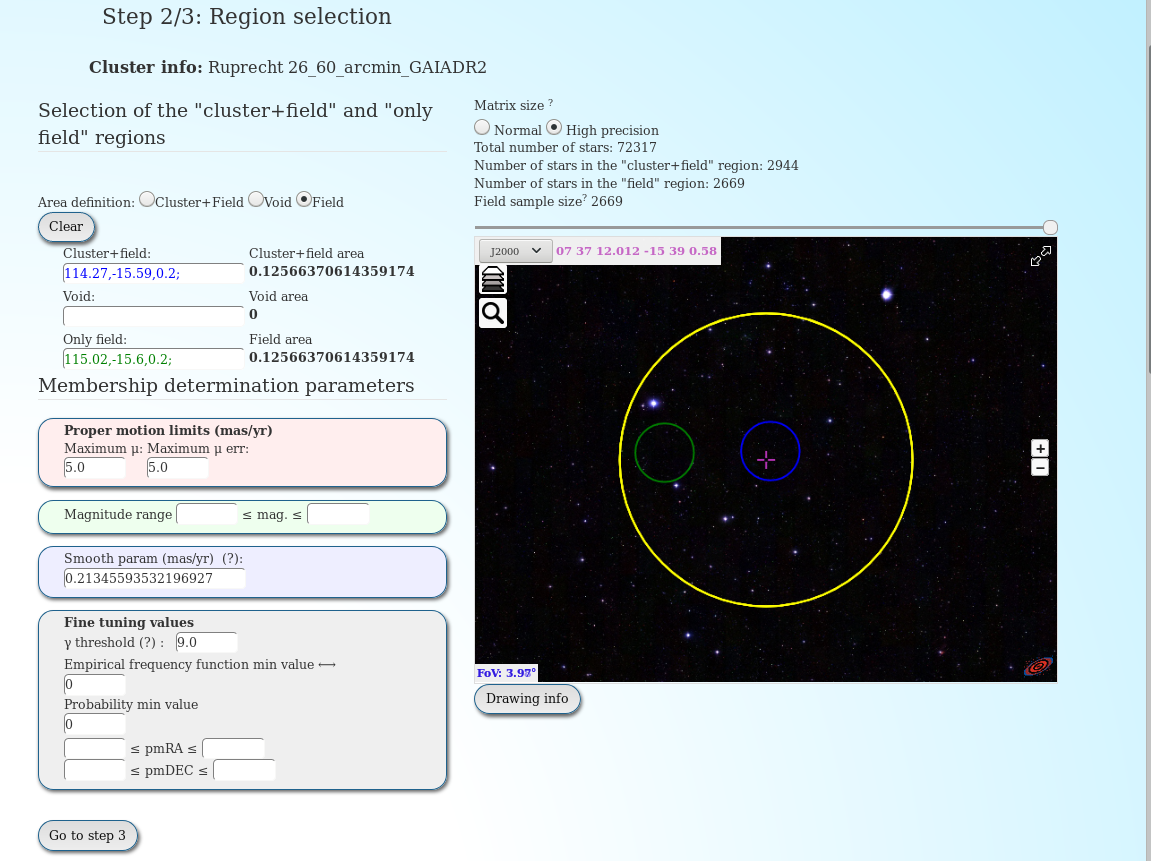}
    \caption{Step~2 where all the fine tuning parameters of the \texttt{Clusterix} procedure are set for Ruprecht~26 and Clusterix~1.}
    \label{fig:Rup26_2a}
\end{figure}

\begin{figure}
	\includegraphics[width=\columnwidth]{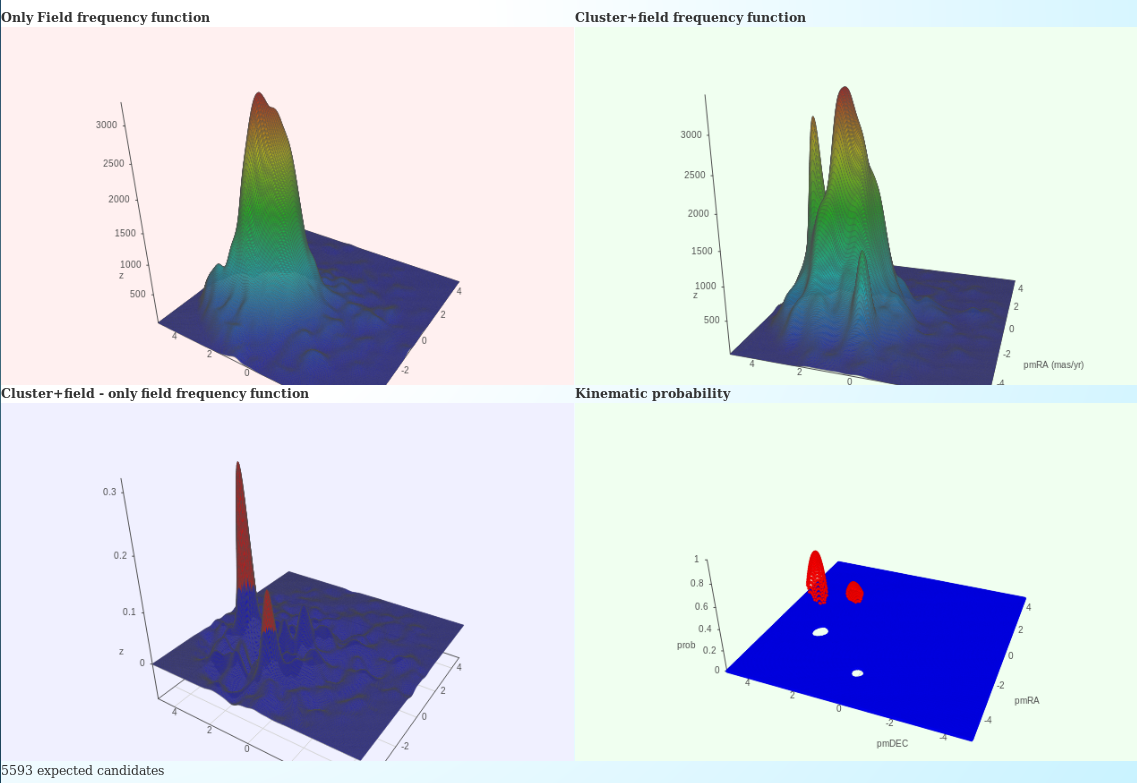}
    \caption{Results of Step~2  \texttt{Clusterix} procedure for Ruprecht~26 and Clusterix~1.}
    \label{fig:Rup26_2b}
\end{figure}

\begin{figure}
	\includegraphics[width=\columnwidth]{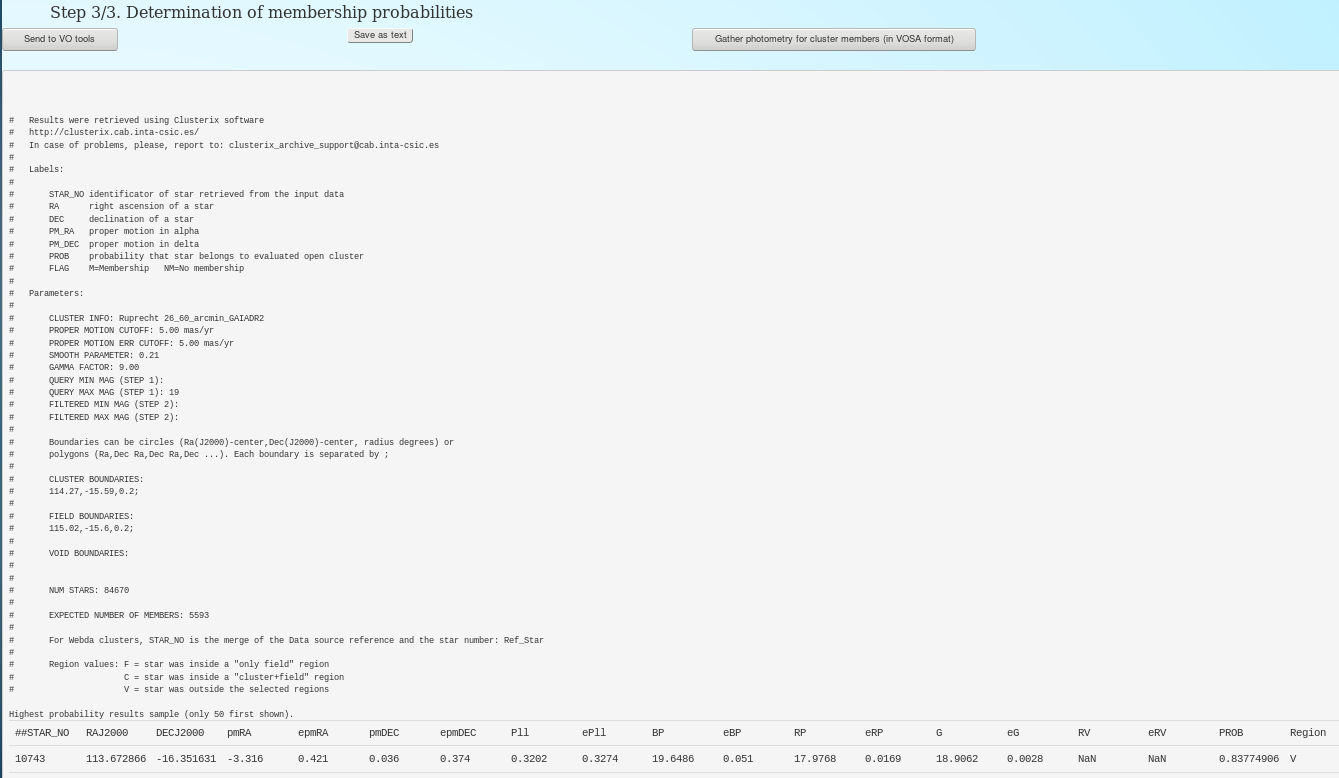}
    \caption{Step~3 of the \texttt{Clusterix} procedure for Ruprecht~26 and Clusterix~1.}
    \label{fig:Rup26_3}
\end{figure}

For Ruprect~26 and Clusterix~1 see Figs.~\ref{fig:Rup26_1},~\ref{fig:Rup26_2a},~\ref{fig:Rup26_2b}, \ref{fig:Rup26_3}.


\bsp	
\label{lastpage}
\end{document}